\documentclass{aastex63}
\usepackage[caption=false]{subfig}
\usepackage{graphicx}
\usepackage{longtable}
\usepackage{multirow}
\usepackage{booktabs}
\usepackage[T1]{fontenc}
\usepackage{appendix}
\graphicspath{{./}{figures/}}
\newcommand{\ra}[4]{$#1^{\rm h}#2^{\rm m}#3^{\rm s}.#4$}
\newcommand{\dec}[4]{$#1^{\circ}2'#3''.#4$}

\published{in ApJ on June 23, 2021} 
\begin{document}

\title{The Study of Dust Formation of Four Type Ibn Supernovae}

\correspondingauthor{Shan-Qin Wang; En-Wei Liang}

\email{shanqinwang@gxu.edu.cn;lew@gxu.edu.cn}

\author{Wen-Pei Gan}
\affiliation{Guangxi Key Laboratory for Relativistic Astrophysics,
School of Physical Science and Technology, Guangxi University, Nanning 530004,
China}

\author{Shan-Qin Wang}
\affiliation{Guangxi Key Laboratory for Relativistic Astrophysics,
School of Physical Science and Technology, Guangxi University, Nanning 530004,
China}

\author{En-Wei Liang}
\affiliation{Guangxi Key Laboratory for Relativistic Astrophysics,
School of Physical Science and Technology, Guangxi University, Nanning 530004,
China}

\begin{abstract}

In this paper, we investigate the early-time optical$-$near-infrared (NIR) spectral energy
distributions (SEDs) of four Type Ibn supernovae (SNe). We find that the SEDs of SN~2010al,
LSQ13ddu, and SN~2015G can be well explained by the single-component blackbody model, while
the SEDs of OGLE-2012-SN-006 cannot. We invoke the double-component model
assuming that the SEDs were produced by the SN photosphere and the heated
dust to fit the optical$-$NIR SEDs of the four SNe Ibn, finding that the derived
temperatures of the dust associated with OGLE-2012-SN-006 favor the scenario that the dust
consists of the graphite grains, and the mass and temperature of dust are
$\sim$$0.5-2.0\times10^{-3}~M_\odot$ and $\sim$ $1200-1300$ K, respectively.
Moreover, our fits for SN~2010al, LSQ13ddu, and SN~2015G show that the upper
limits of the masses of the dust associated with the three SNe Ibn are respectively
$1.45\times 10^{-5}~M_\odot$, $5.9\times 10^{-7}~M_\odot$, and $2.4\times 10^{-7}~M_\odot$.
A further analysis demonstrates that the inferred radius of the dust shell
surrounding OGLE-2012-SN-006 is significantly larger than that of the SN ejecta
at early epochs, indicating that the NIR excesses of the SEDs of OGLE-2012-SN-006 were
produced by a preexisting dust shell. Our study for the early-time SEDs of
four SNe Ibn, together with the previous studies
and the fact that some SNe showed the evidence of dust formation at the late-time SEDs,
indicates that at least $\sim$1/3 of SNe Ibn show evidence for dust formation.

\end{abstract}


\section{Introduction}
\label{sec:intro}

Type Ibn supernovae (SNe Ibn; \citealt{Pas2016,Hoss2017}) are believed to be
produced by a massive star surrounded by a dense helium-rich circumstellar material (CSM)
that can be either winds blown by the SN progenitors or shells expelled
by the progenitors \citep{Fol2007,Pas2007}. In this scenario, the interaction between the SN ejecta and the CSM
produces forward shocks and reverse shocks. The forward shocks ionized the helium-rich CSM, which can emit
narrow- and/or intermediate-width emission lines when the ionized helium recombine.

Assuming the ratio of the dust mass to the gas mass is 1:100 which is a standard value (e.g., \citealt{Tartaglia2020}),
there must be a moderate mass of dust surrounding the progenitor of an SN Ibn before the SN explosion.
Alternatively, the ejecta$-$CSM interaction might produce new dust in the shocked regions.
The UV and optical photons from the SNe would heat the preexisting
and/or new dust, then the heated dust would yield excess emission peaking at
near infrared (NIR) or middle-IR (MIR).

Therefore, the IR excesses at the early epochs and/or the IR detections
at the (very) late epochs can provide useful information for
exploring the dust formation associated with an SN.

To date, at least a few dozen SNe have been confirmed to be surrounded by
a moderate amount ($\sim$$10^{-6}-10^{-2}~M_\odot$; see, e.g., \citealt{Fox2011,Fox2013,Szalai2019})
of dust. For comparison, only one SN Ibn (SN~2006jc) was previously found to have
evidence of dust formation \citep{Mattila2008,Smith2008}.

It is an interesting issue to search for the evidence of dust formation
associated with other SNe Ibn. One of prevailing methods confirming the dust
formation associated with SNe is studying the optical and NIR/MIR
spectral energy distributions (SEDs) of the SNe since the dust emission
would yield IR excesses.

So far, only six SNe Ibn (SN~2010al, \citealt{Pas2015a}; OGLE-2012-SN-006, \citealt{Pas2015b};
SN~2006jc, \citealt{Fol2007,Smith2008,Pas2008,Tomi2008}; PS1-12skm, \citealt{Sanders2013,Hoss2019};
LSQ13ddu, \citealt{Clark2020}; SN~2015G \citealt{Shiv2017,Hoss2017})
have been observed in both optical and NIR ($JHK$) bands.
Among these six SNe Ibn, as mentioned above, SN~2006jc
was previously found to have SEDs
showing NIR excesses from dust that had been interpreted by
a two-component model assuming that the NIR excesses were produced by the preexisting dust
\citep{Dwek1985} in the CSM and/or the newly formed dust in the shocked CSM \citep{Mattila2008,Smith2008}.
In contrast, \cite{Sanders2013} point out that the SEDs of PS1-12sk do not show significant NIR excess
(see their Figure 4).

The evidence for dust formation in the remaining four
SNe Ibn with optical and NIR data had not been studied.
In this paper, we study the SEDs of the four SNe Ibn
(SN~2010al, OGLE-2012-SN-006, LSQ13ddu, and SN~2015G)
that have $JHK$ photometry but have not been studied in detail.
In section \ref{sec:SED},
we use the single-component model
and double-component model to fit the SEDs of the four SNe Ibn.
We discuss our results and draw conclusions in
Sections \ref{sec:discussion} and \ref{sec:Con}, respectively.

\section{The Optical and NIR SEDs and the SED fitting}
\label{sec:SED}

In this section, we study the SEDs of four SNe Ibn
(SN~2010al, OGLE-2012-SN-006, LSQ13ddu, and SN~2015G)
whose details are listed in Table \ref{table:details}.
The redshifts ($z$) of the SNe in the sample are between 0.005 and 0.06,
and the observations for the sample cover both the optical and the NIR bands.

First, we fit the SEDs of the SNe Ibn
using the single-component model, which assumes that the SEDs are produced
by the blackbody emission from the SN photosphere,
\begin{eqnarray}
F_{\nu,{\rm ph}} = {\pi}B_\nu(T_{\rm ph})\frac{R_{\rm ph}^2}{D_L^2},
\label{eqn:flux-ph}
\end{eqnarray}
where $F_{\nu,{\rm ph}}$ and $B_\nu(T_{\rm ph}) = (2 h{\nu}^3/c^2)(e^{\frac{h{\nu}}{k_{\rm b}T_{\rm ph}}}-1)^{-1}$
are the flux and the intensity of the blackbody radiation of the SN photosphere, respectively.
$R_{\rm ph}$, $T_{\rm ph}$, and $D_L$ are the radius, temperature of the
SN photosphere, and the luminosity distance of the SN, respectively.

\begin{figure}[tbph]
\vspace{80pt}
\begin{center}
\includegraphics[width=0.45\textwidth,angle=0]{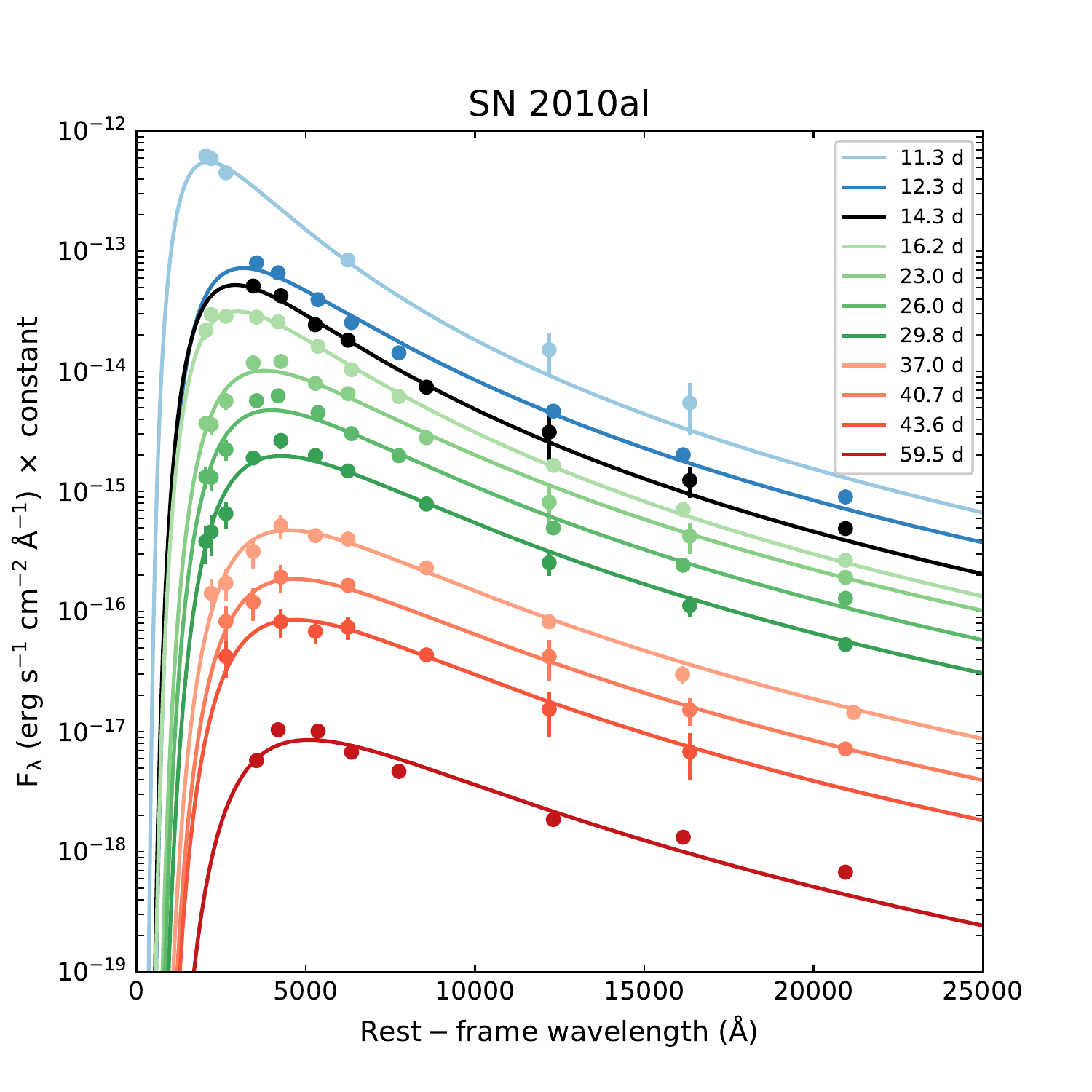}
\includegraphics[width=0.45\textwidth,angle=0]{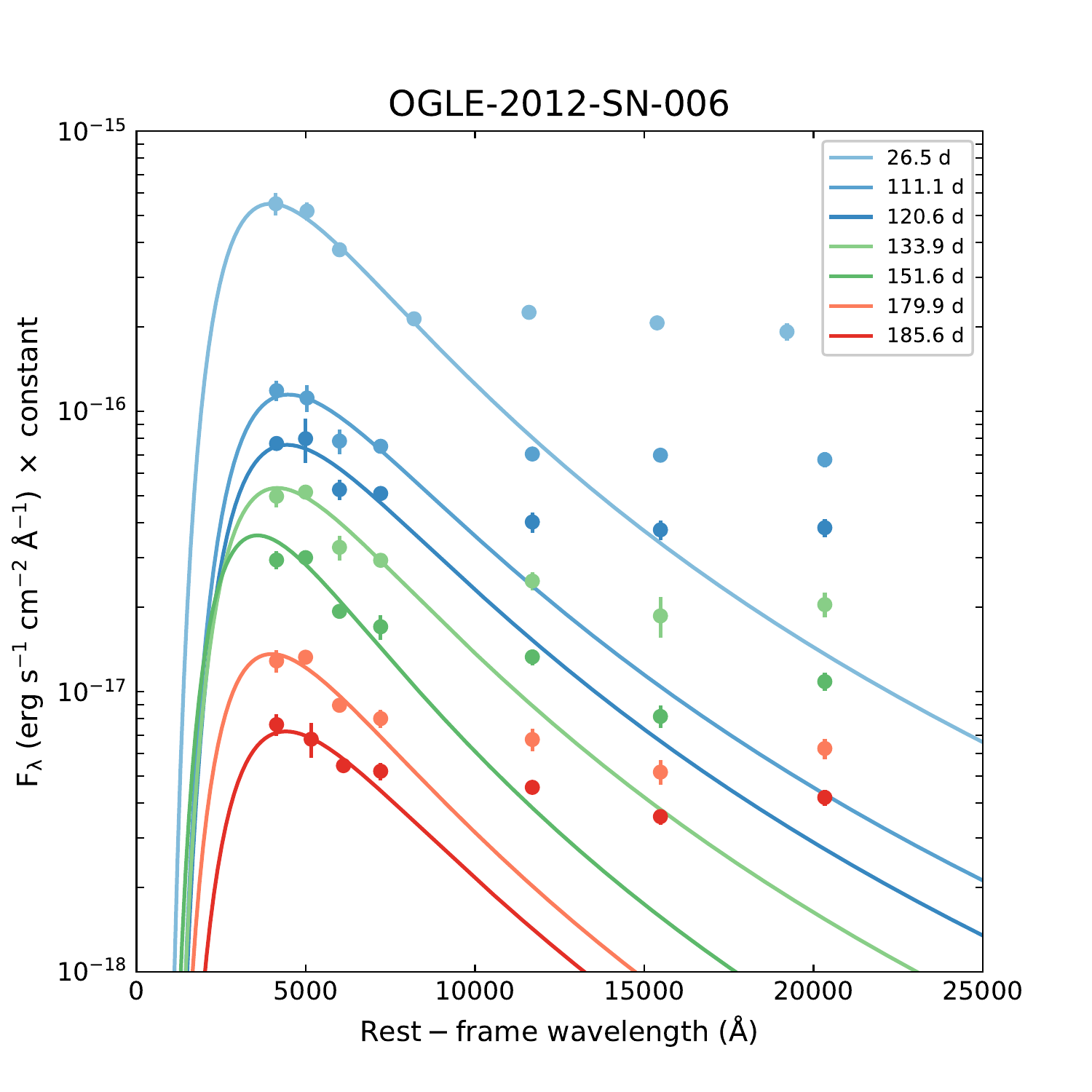}
\includegraphics[width=0.45\textwidth,angle=0]{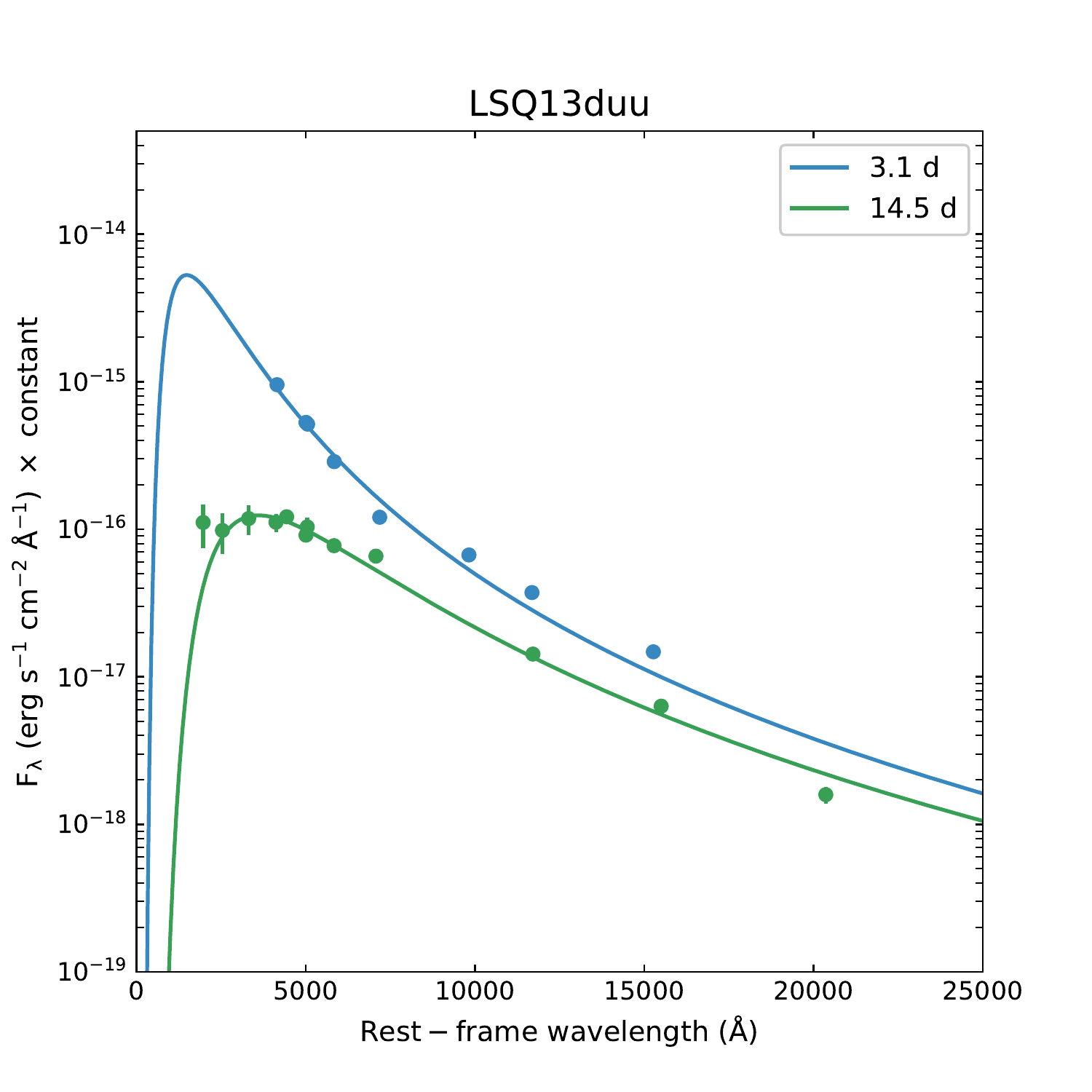}
\includegraphics[width=0.45\textwidth,angle=0]{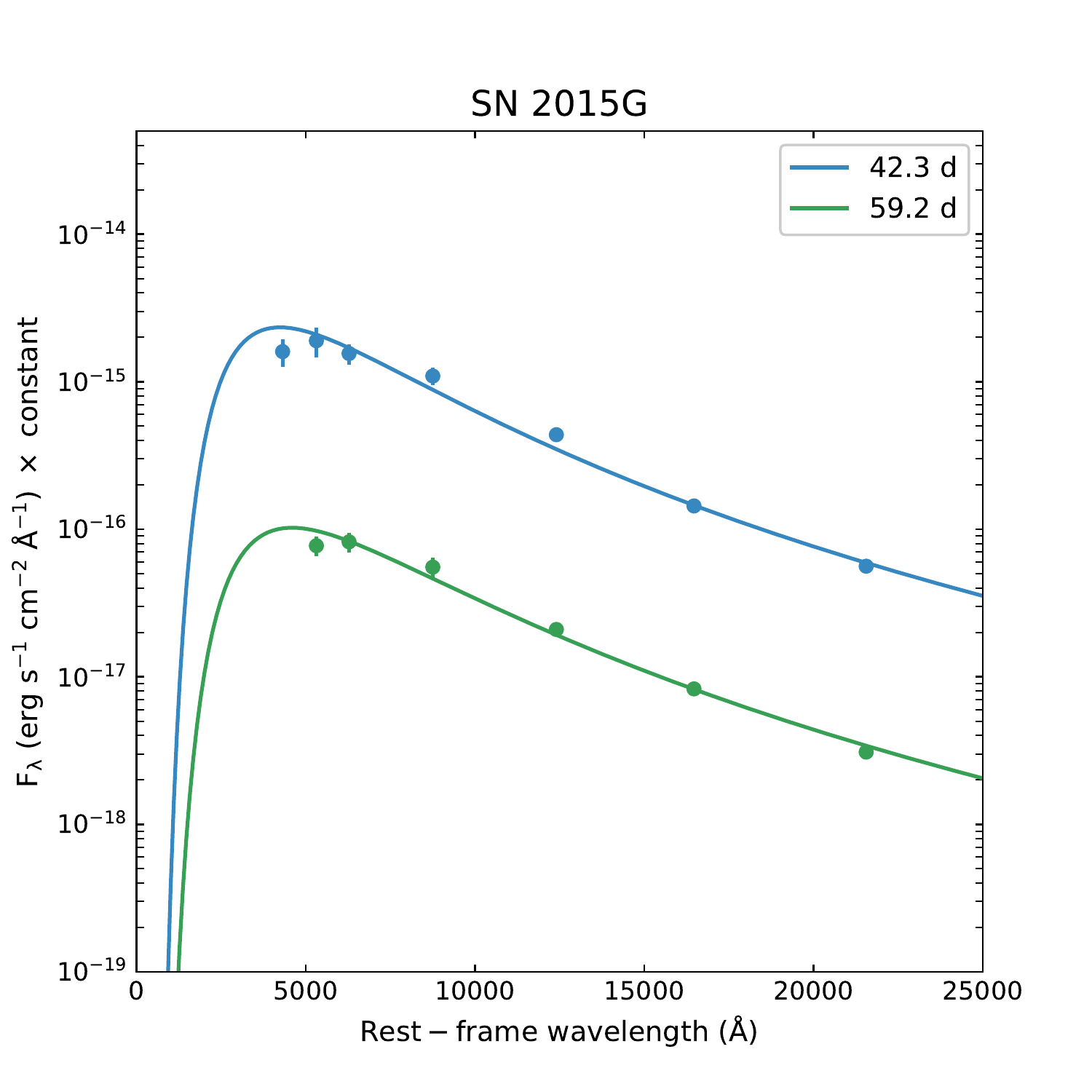}
\end{center}
\caption{The optical and NIR SEDs of SN~2010al, OGLE-2012-SN-006, LSQ13ddu, and SN~2015G
and the fits of the single-component blackbody model. The data are from
the references listed in Table \ref{table:details}. For clarity, the flux at all epochs are shifted
by adding different constants.}
\label{fig:SED1}
\end{figure}

To get best-fitting parameters and the 1$\sigma$ range of the parameters,
we adopt the Markov Chain Monte Carlo (MCMC) method
by using the \texttt{emcee} Python package \citep{Foreman-Mackey2013}.
The best-fitting theoretical SEDs of these four SNe Ibn are
shown in Figure \ref{fig:SED1}.

It can be found that the SEDs of SN~2010al, LSQ13ddu, and SN~2015G
can be well explained by the single-component blackbody model, indicating
that there is no significant evidence of the dust formation associated with
these three SNe.

In contrast, the SEDs of OGLE-2012-SN-006 cannot be explained by this
model since the SEDs show evident NIR ($JHK$) excesses.
To account for the NIR ($JHK$) excesses of the SEDs of
OGLE-2012-SN-006 and constrain the
upper limits of the dust masses associated with SN~2010al, LSQ13ddu, and SN~2015G,
a two-component model is needed.
This model assumes a portion of the UV and optical flux
produced in the ejecta of SNe heats the dust shells and the
heated dust emits IR photons, producing the IR excesses \citep{Dwek1983}.
Hence, we employ the double-component model assuming
that the flux is the sum of the emission of SN photosphere and the dust shell.

The total flux of the two-component model can be written as
\begin{eqnarray}
F_{\nu}=F_{\nu,{\rm ph}}+F_{\nu,{\rm d}}
\end{eqnarray}
where the first component is from the SN photosphere
and described by Equation (\ref{eqn:flux-ph}),
while the second component, which is from the heated dust, can be expressed as
\citep{Hildebrand1983}
\begin{eqnarray}
F_{\nu,{\rm d}} = B_\nu(T_{\rm d})\frac{M_{\rm d} \kappa_\nu(a)}{D_L^2},
\label{eqn:flux-dust}
\end{eqnarray}
Here, $F_{\nu,{\rm d}}$ is the flux from the dust shell,
$B_\nu(T_{\rm d})= (2 h{\nu}^3/c^2)(e^{\frac{h{\nu}}{k_{\rm b}T_{\rm d}}}-1)^{-1}$ is the
intensity of the black-body radiation of a dust grain, and
$M_{\rm d}$ is the mass of the dust shell.

The mass absorption coefficient ($\kappa_\nu(a)$) of the dust grains
plays a crucial role in the model.
Following \cite{Dwek1983}, $\kappa_\nu(a)$ can be written as
\begin{equation}
\label{eqn:kappa}
\kappa_\nu (a) = \frac{3Q_\nu(a)}{4 \rho a},
\end{equation}
where $\rho$ and $a$ are the mass density and radius of the spherical dust grain, respectively.
Fixing the values of $\rho$ and $a$, $\kappa_\nu(a)$ is uniquely determined by the value of
the emission efficiency $Q_\nu(a)$,
which can be calculated by Mie theory for the dust grains.
As suggested by previous studies (e.g., \citealt{Laor1993,Fox2010,Fox2011,Stritzinger2012}),
the grains in the dust shell surrounding SNe can be either graphite or silicate
whose mass densities are $2.26~\rm g~cm^{-3}$ and $3.3~\rm g~cm^{-3}$, respectively \citep{Laor1993}.

\begin{figure}[tbph]
\vspace{140pt}
\begin{center}
\includegraphics[width=0.45\textwidth,angle=0]{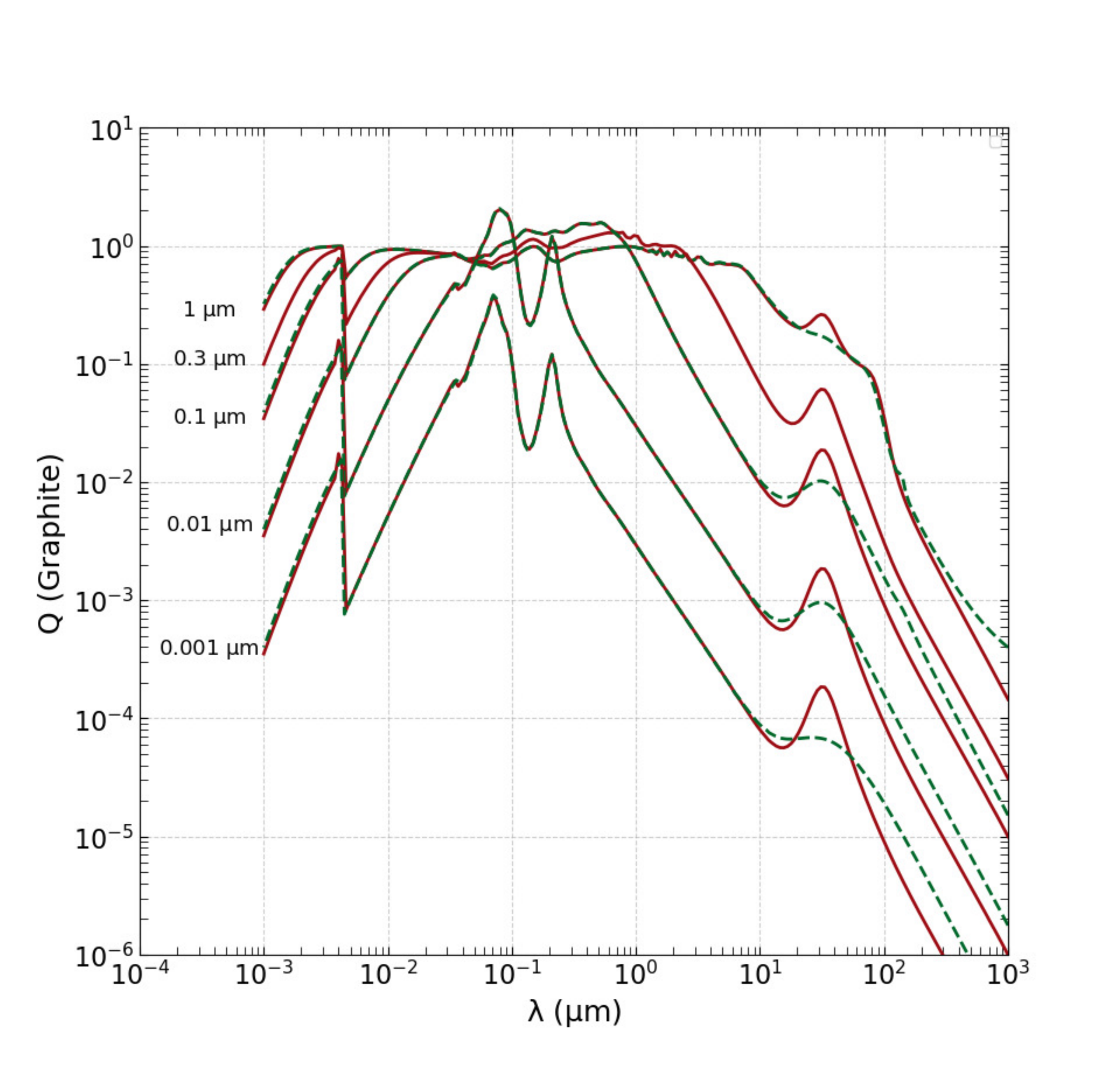}
\includegraphics[width=0.45\textwidth,angle=0]{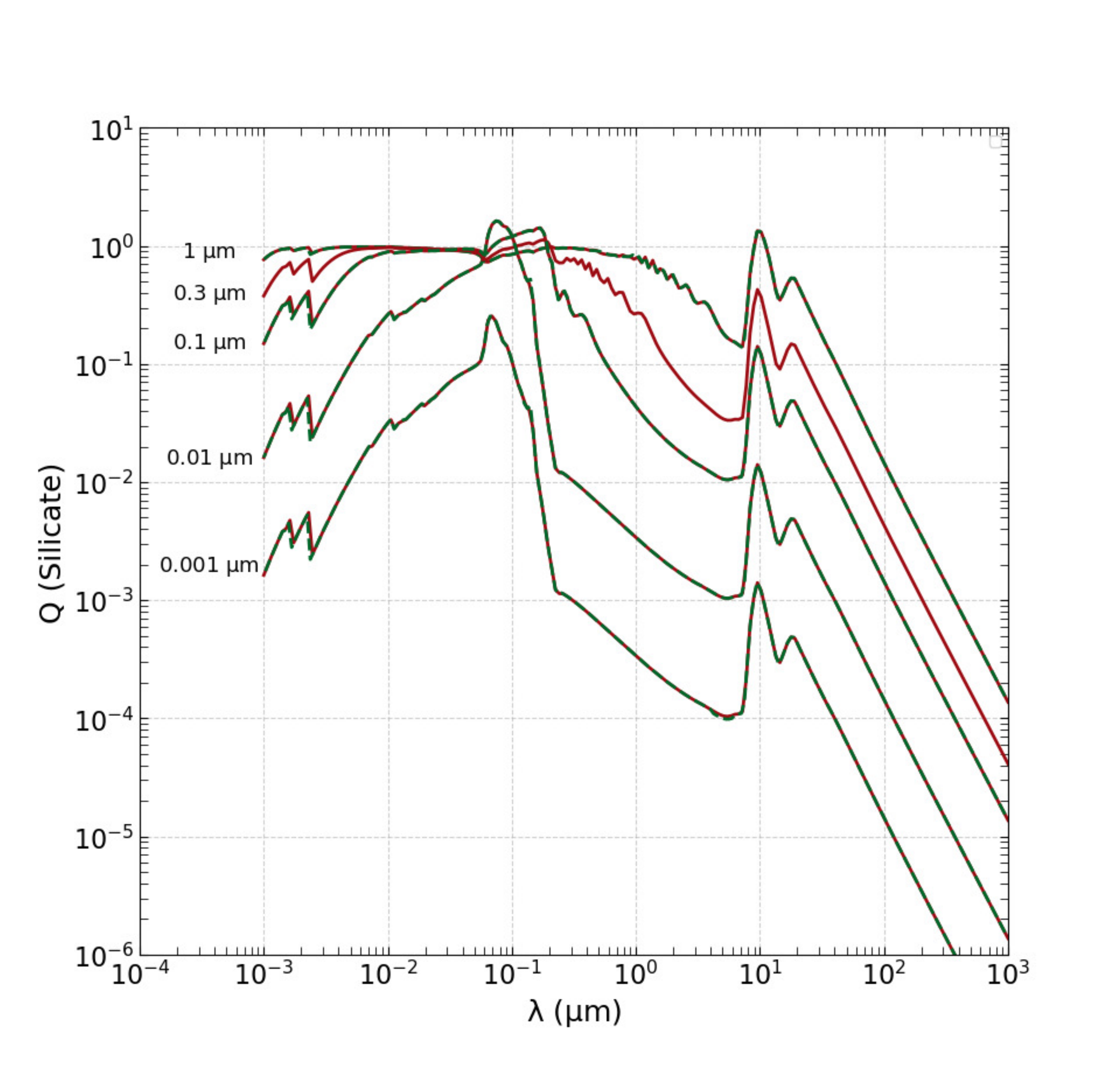}
\end{center}
\caption{The emission efficiencies ($Q_\nu(a)$) for graphite (left panel) and silicate
(right panel) for $a$=0.001, 0.01, 0.10, 0.3, and 1.0
$\mu$m. The solid lines present the curves we calculated using Mie theory,
while the dashed lines present the curves yielded by the data obtained from
Draine's website. Note that the dashed line corresponding the case of
$a$ = 0.3 $\mu$m is absent since the website provide no data for this case.}
\label{fig:Q}
\end{figure}

Using Mie theory, we plot in Figure \ref{fig:Q} the emission efficiencies ($Q_\nu(a)$) for both graphite and silicate.
The data of the refractive index and the extinction coefficient of graphite
and silicate needed for our calculations are from \citet{Draine2003} which are presented in Draine's website.
\footnote{https://www.astro.princeton.edu/~draine/dust/dust.diel.html}
For comparison, we overplot the emission efficiency curves using the data downloaded from
Draine's website where the data were yielded by using the equations listed in
\citet{Draine1984} and \citet{Laor1993}.
\footnote{Note that similar curves produced by Mie theory were shown in
Figure $4$ of \cite{Fox2010} for $a$=0.001, 0.01, 0.10, and 1.0
$\mu$m at the wavelength range between $10^{-2}$ and $\sim$10 $\mu$m.}
For silicate, the two groups of curves are consistent with each other.
For graphite, the two groups of curves are slightly different if
$\lambda$ is $\lesssim$$5\times 10^{-3}$ $\mu$m or $\gtrsim$10 $\mu$m.
The discrepancy is due to the fact that the curves of $Q_\nu(a)$ from Draine's website
is based on the refractive index and the extinction coefficient from
\citet{Laor1993} while the curves of $Q_\nu(a)$ we calculate are based on
the refractive index and the extinction coefficient from
\citet{Draine2003}. Nevertheless, the discrepancy is irrelevant since the range of
the effective wavelength of the NIR bands ($JHK$) is $\sim$$1.2-2.2$ $\mu$m.

In principle, the radius of the dust grains $a$ can be included in the
MCMC; however, the fitting for SEDs of different epochs might
result in different $a$ while $a$
ought to be a constant at different epochs for the same SN.
Here, we assume that $a=0.10$ $\mu$m.
So the free parameters of the double-component SED fitting are
$R_{\rm ph}$, $T_{\rm ph}$, $T_{\rm d}$, and $M_{\rm d}$.

\begin{figure}[tbph]
\vspace{80pt}
\begin{center}
\includegraphics[width=0.45\textwidth,angle=0]{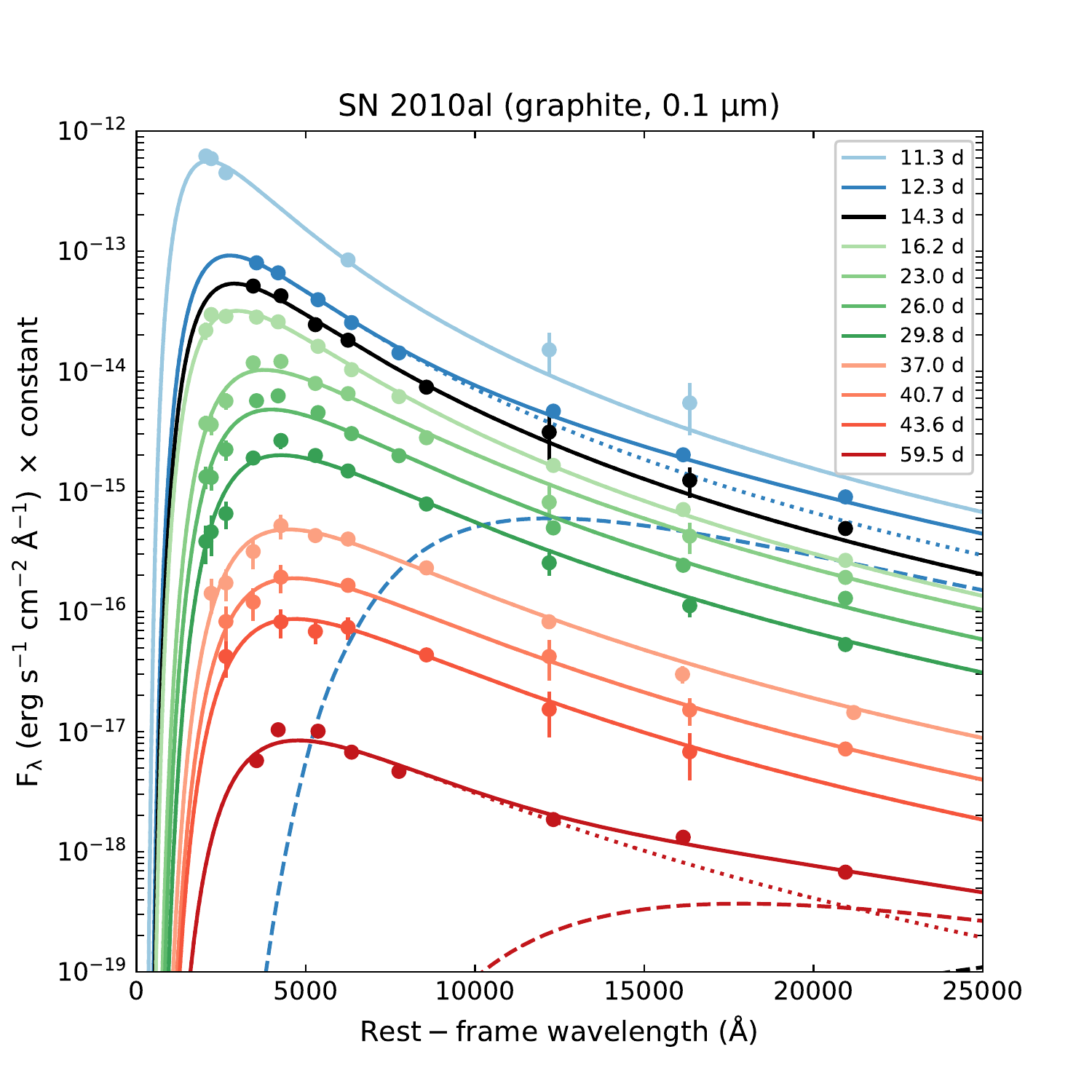}
\includegraphics[width=0.45\textwidth,angle=0]{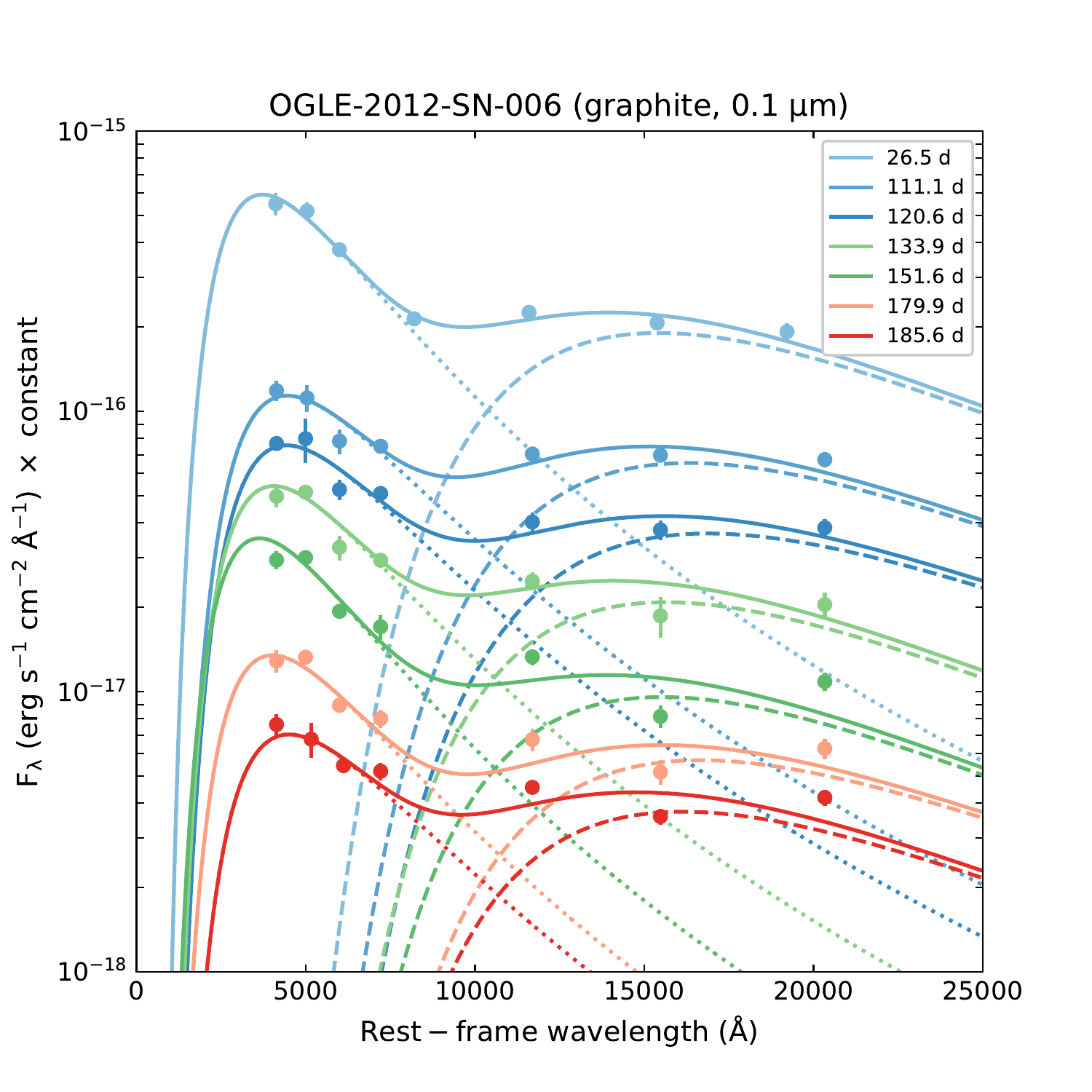}
\includegraphics[width=0.45\textwidth,angle=0]{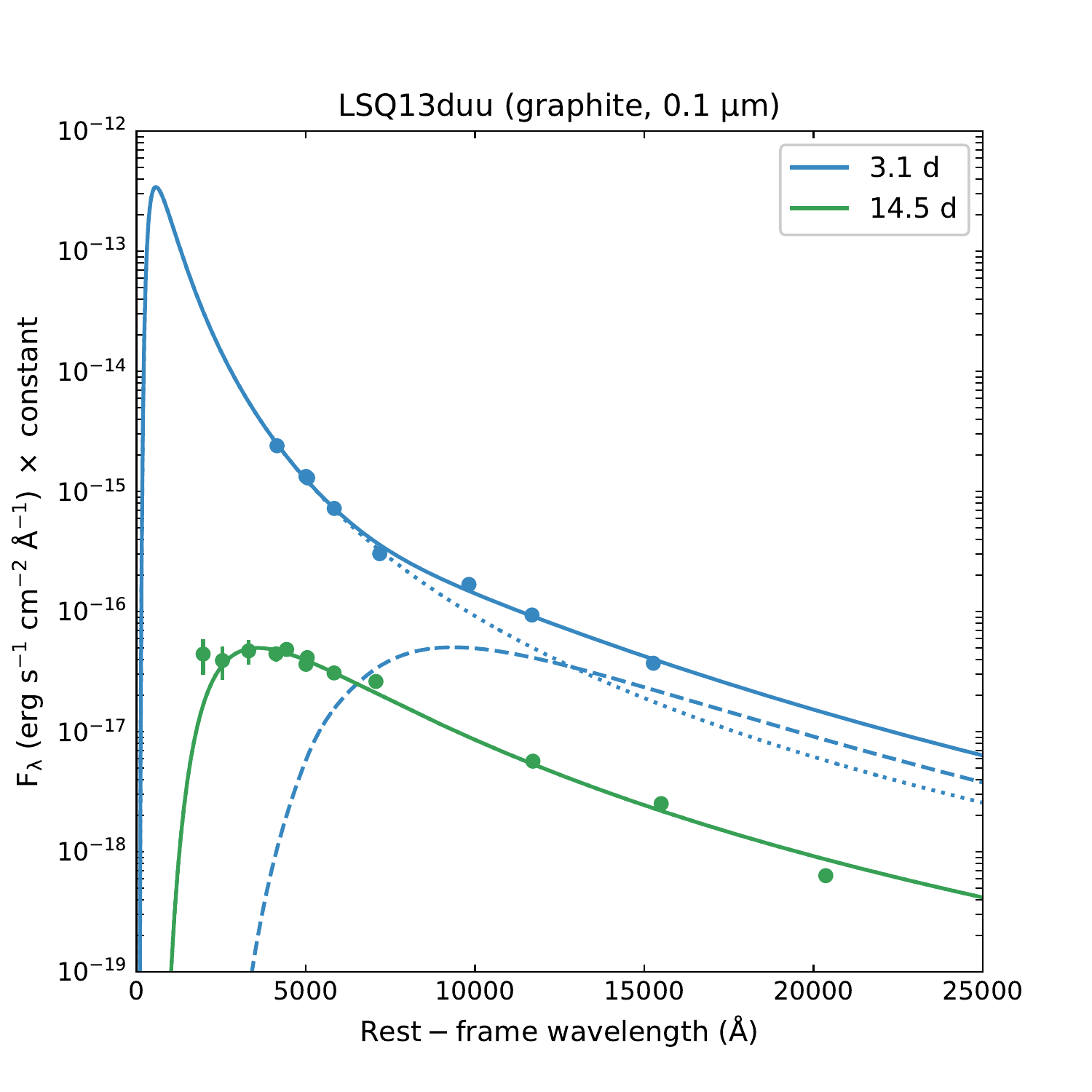}
\includegraphics[width=0.45\textwidth,angle=0]{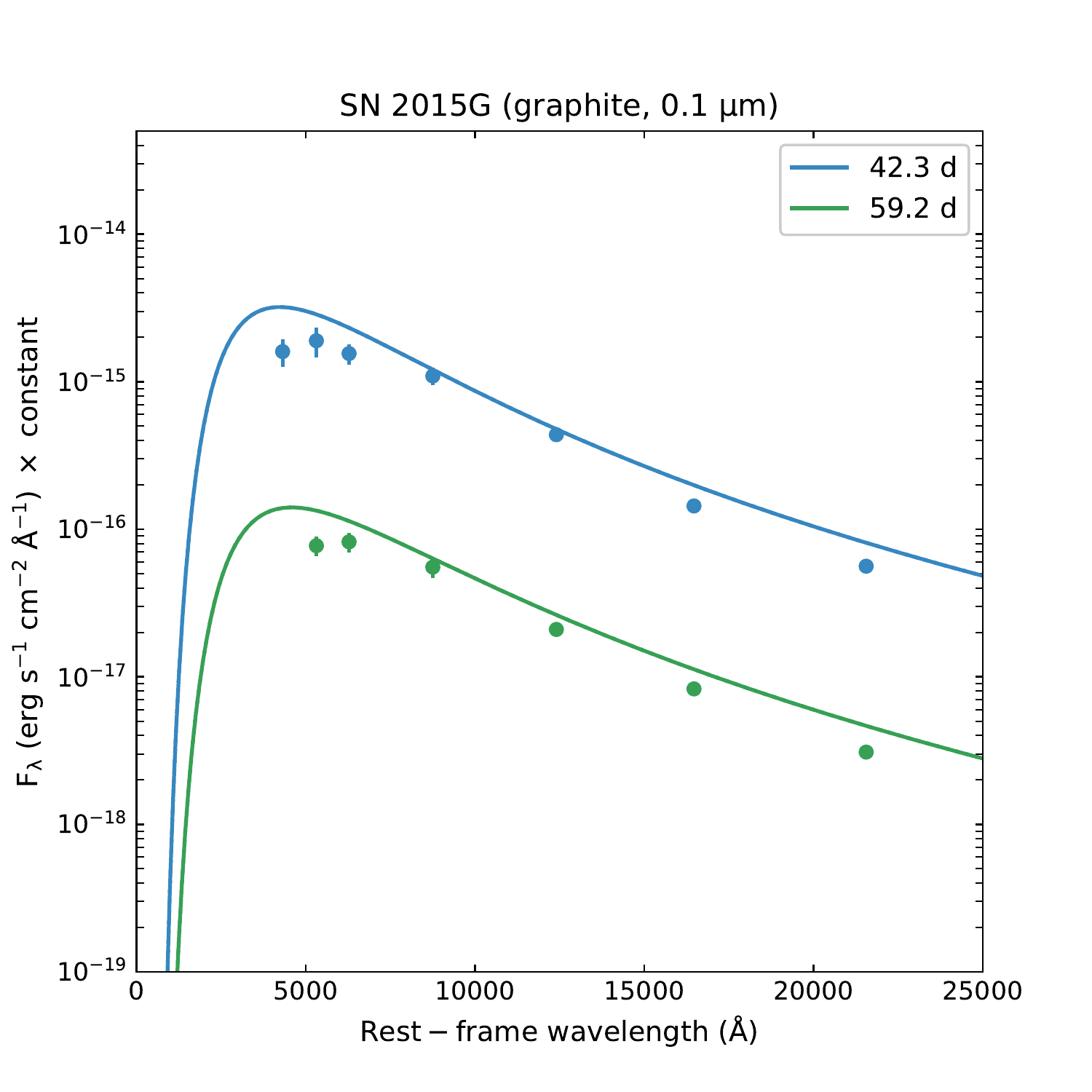}
\end{center}
\caption{The best-fitting SEDs of SN~2010al, OGLE-2012-SN-006, LSQ13ddu, and SN~2015G obtained by using the
two-component model. The dotted, the dashed, and the solid lines present the flux of SN photospheres, dust,
and the sum of the two components, respectively.
The data are from the references listed in Table \ref{table:details}. For clarity, the plots at all epochs have been shifted
vertically.}
\label{fig:SED-double}
\end{figure}

The SEDs of the four SNe Ibn at all epochs can be well explained
by the double-component model (see Figure \ref{fig:SED-double});
the corresponding corner plots (for graphite dust) are shown in
Figures \ref{fig:2010corner},
\ref{fig:2012corner}, \ref{fig:2013corner}, and \ref{fig:2015corner}.
The best-fitting parameters of the two-component model are presented in
Table \ref{table:SED_PARAM-double}.
We note that the model fits for late-time SEDs are not
very well. This is because the simple double-component
model must yield smooth theoretical SEDs that cannot fully
match the observed SEDs that have undulation features.
Nevertheless, the slight deviations between the theoretical
two-component SEDs and the observed ones are within a reasonable range.

The derived temperatures can be used to judge the composition
of the dust. The evaporation temperatures of silicate and
graphite are $\sim$1100-1500 K (e.g., \citealt{Laor1993,Mattila2008,Gall2014})
and $\sim$1900 K (e.g., \citealt{Stritzinger2012}), respectively.
Comparing these two values to the best-fitting temperatures
of the silicate and graphite, we find the former can be excluded and
the latter is favored.

For OGLE-2012-SN-006, the best-fitting values of $T_{\rm ph}$,
$R_{\rm ph}$, $T_{\rm d}$ and $M_{\rm d}$ at different epochs are
$\sim$$6400-7800$ K, $\sim$$0.9-1.9 \times 10^{15}$ cm,
$\sim$$1200-1300$ K, and $\sim$$0.5-2.0 \times 10^{-3}~M_\odot$,
respectively. It should be noted that the inferred mass of the dust is
just a lower limit since the dust grains with lower temperatures would
produce emission peaks at longer wavelengths and the $JHK$ flux from the
cold dust might be lower than the detection limit.

The fit for the SED of LSQ13ddu at day 3.1 can be excluded
since the derived temperatures for both two cases (2262 K and 2599 K)
exceed 1900 K. The best-fit values of the masses derived from the SEDs of SN~2010al,
the rest one SED of LSQ13ddu, and the SEDs of SN~2015G are $1.45\times 10^{-5}~M_\odot$,
$5.9\times 10^{-7}~M_\odot$, and $2.4\times 10^{-7}~M_\odot$, respectively.
The derived values of the masses of the dust associated with LSQ13ddu and SN~2015G are just the upper limits.
While the possibility that there is a dust shell surrounding SN~2010al cannot
be excluded, the derived mass of the dust associated with SN~2010al can also be regarded as an upper limit since,
as mentioned above, the slight deviation from the data might be due to the fact that
the simple blackbody model must
produce smooth SEDs and the observed SEDs have undulation features.
Hereafter, we only study the properties of the dust of OGLE-2012-SN-006.

\section{Discussion}
\label{sec:discussion}

\subsection{Preexisting or New dust?}

The NIR excesses of the SEDs can be produced by preexisting and/or new dust.
To determine the nature of the dust of OGLE-2012-SN-006, a further analysis is needed.
We first derived the luminosities of the SN photosphere ($L_{\rm ph}$) and the dust
($L_{\rm d}$), as well as the radius of the dust shell ($R_{\rm d}$),
based on the parameters listed in Table \ref{table:SED_L}.

The luminosities of the SN photospheres and the dust shells can be respectively calculated by the equations
\begin{eqnarray}
L_{\rm ph} = 4 \pi \sigma T_{\rm ph}^4R_{\rm ph}^2,
\label{eqn:lum-dust}
\end{eqnarray}
and
\begin{eqnarray}
L_{\rm d} = 4\pi M_{\rm d} \int_0^{\infty} B_\nu(T_{\rm d}) \kappa_\nu(a)d\nu,
\label{eqn:lum-dust}
\end{eqnarray}
where $\sigma = 5.67 \times 10^{-5} \,\mathrm{erg~cm^{-2}~K^{-4}}$ is the Stefan Boltzmann constant.

The lower limits of $R_{\rm d}$ at different epochs can be derived by the equation
\begin{equation}
R_{\rm bb} = \bigg(\frac{L_{\rm d}}{4 \pi \sigma T_{\rm d}^4}\bigg)^{1/2}.
\label{eqn_rbb}
\end{equation}
As pointed by \cite{Fox2011}, Equation (\ref{eqn_rbb}) can only be
used for an optically thick dust shell. For an optically thin dust
shell, the value derived by Equation (\ref{eqn_rbb}) is the
lower limit of the radius of the dust shell, i.e., $R_{\rm d} > R_{\rm bb}$.
The values of $L_{\rm ph}$, $L_{\rm d}$, and $R_{\rm d}$ of OGLE-2012-SN-006
are presented in Table \ref{table:SED_L}.

By comparing Tables \ref{table:SED_PARAM-double} and \ref{table:SED_L},
we find that the radius of the SN photosphere of OGLE-2012-SN-006
at day 26.5 ($\sim$1.9 $\times 10^{15}$ cm)
is significantly smaller than the radius of the dust shell ($\gtrsim$$6.8\times 10^{16}$ cm)
at the same epoch. At the early-time evolution, the radii of the surfaces of SN ejecta
are slightly larger than that of the SN photospheres and we can assume that the two
are approximately equal to each other.
This result indicates that the radius of the dust shell surrounding OGLE-2012-SN-006 is
significantly larger than that of the
SN ejecta and the NIR excesses might be produced by the preexisting dust.

\subsection{The Evolution of Some Parameters of OGLE-2012-SN-006}

\begin{figure}[tbph]
\vspace{80pt}
\begin{center}
\includegraphics[width=0.45\textwidth,angle=0]{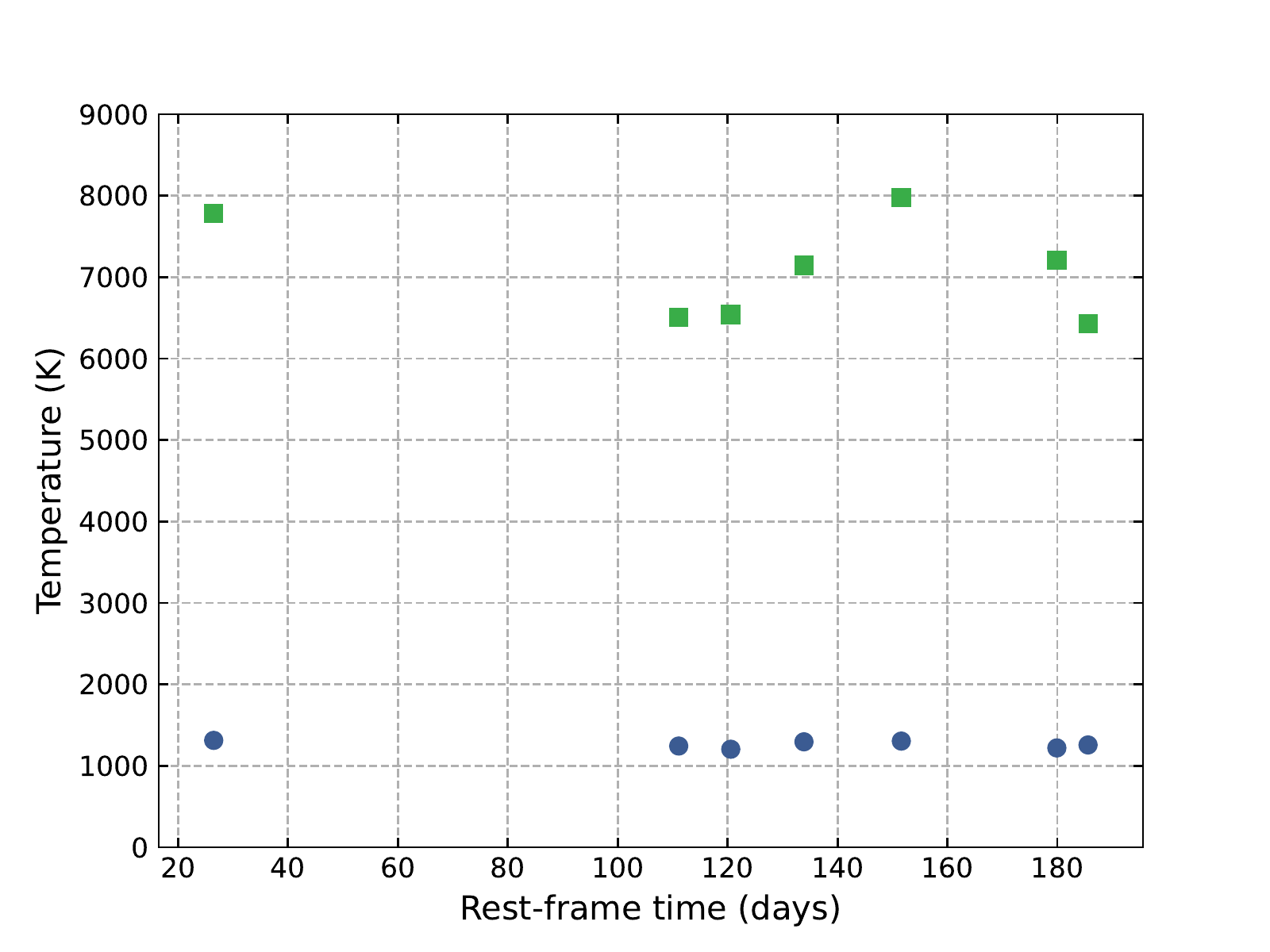}
\includegraphics[width=0.45\textwidth,angle=0]{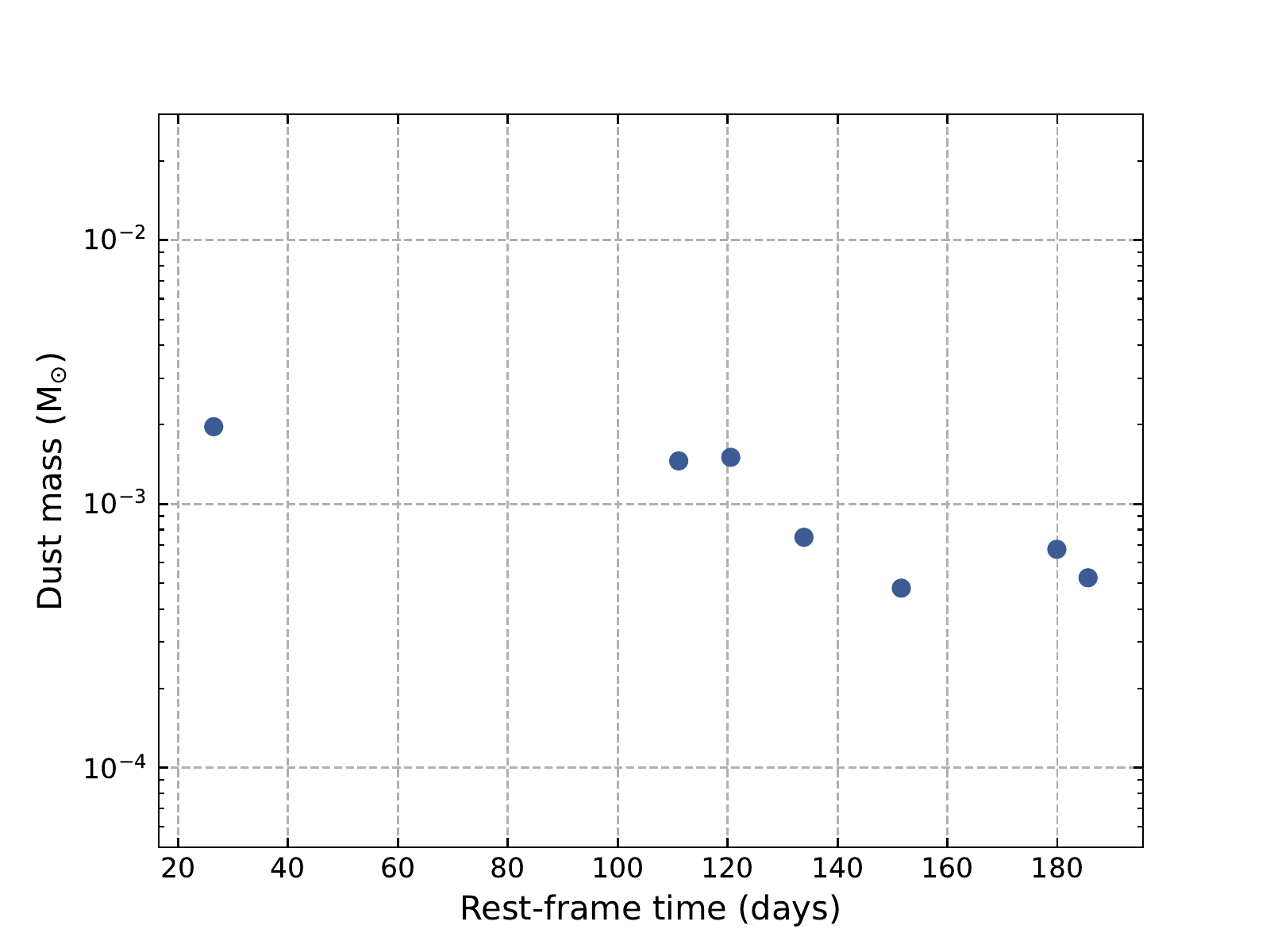}
\includegraphics[width=0.45\textwidth,angle=0]{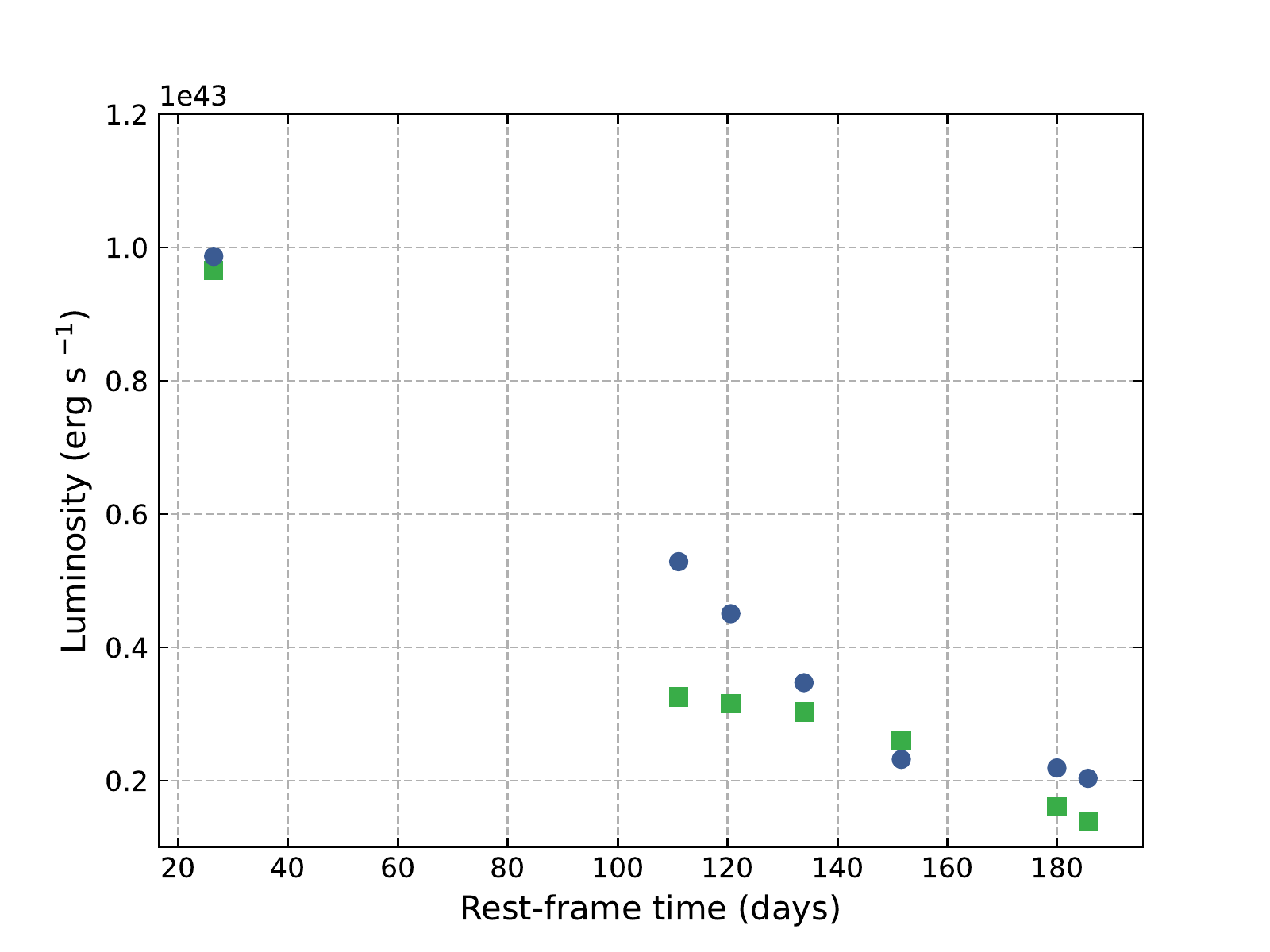}
\includegraphics[width=0.45\textwidth,angle=0]{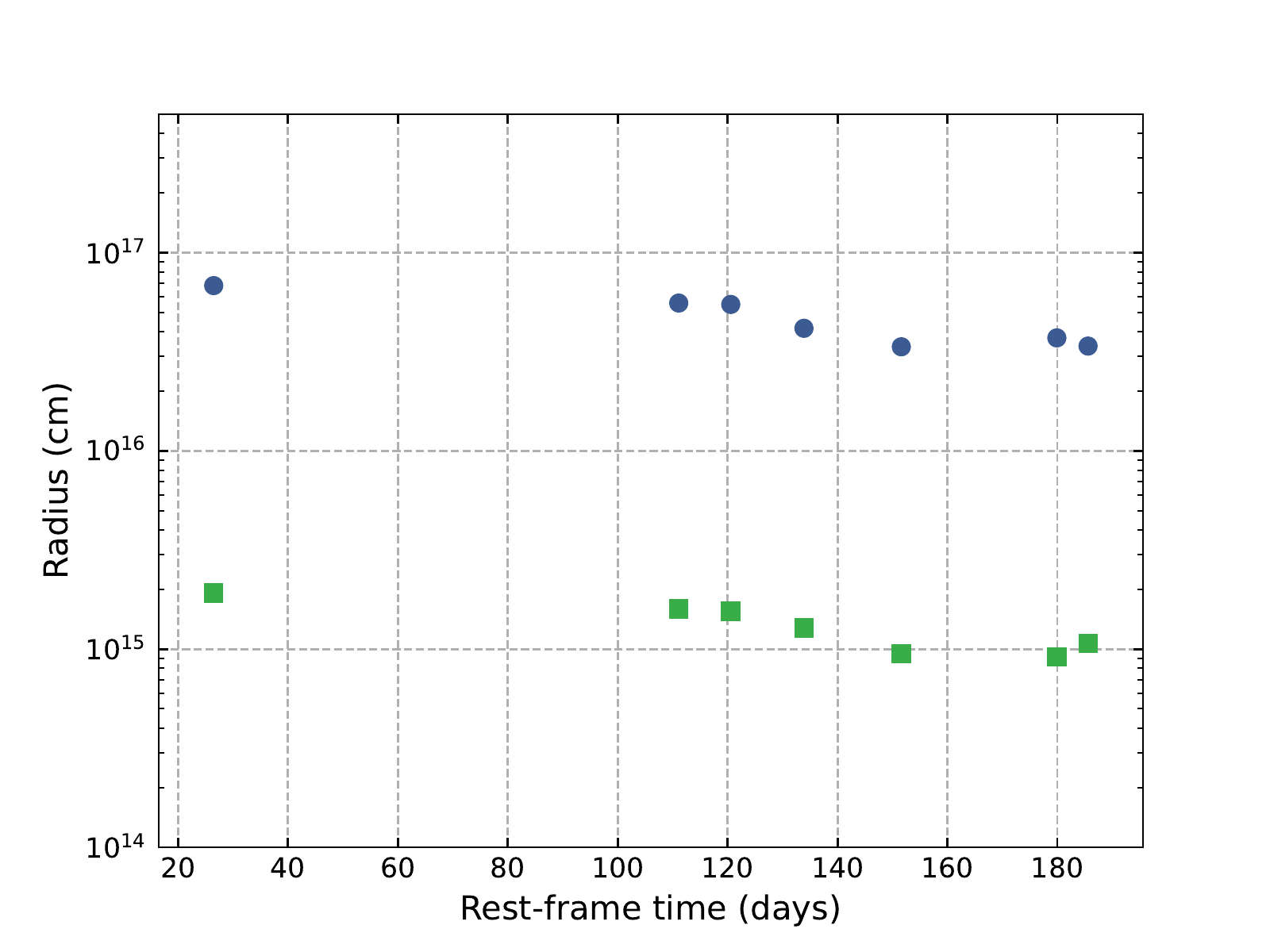}

\end{center}
\caption{The evolution of the temperatures (the top-left panel), the dust masses (the top-right panel),
the luminosities of the SN photosphere and the dust (the bottom-left panel), and the radii (the bottom-right panel)
of OGLE-2012-SN-006.}
\label{fig:evo}
\end{figure}

In Figure \ref{fig:evo}, we plot the
evolution of the temperature, the radius,
and the luminosity of the photosphere of OGLE-2012-SN-006
and the dust as well as the mass of the dust.

Throughout the evolution, the temperature of the SN photosphere
of OGLE-2012-SN-006
is significantly higher than that of the dust, while the
radius of the SN photosphere
is significantly smaller than that of the dust shell;
the luminosity of SN photosphere is lower than
that of the dust shell, except for the epoch of
day 151.6 when the SN photosphere luminosity is
slightly higher than dust shell luminosity.

\subsection{Comparison with Other Interacting SNe}

Comparing the parameters of OGLE-2012-SN-006 with
that of other interacting SNe (SNe Ibn and IIn) would
provide us more information.
\cite{Fox2011} used Spitzer to perform a survey for 68 nearby ($D_{\rm}<250$ Mpc) SNe IIn
discovered between 1999 and 2008 and found that 10 events ($\sim$15\% of the sample)
show late-time IR emission that can be explained by the emission of the dust.
\cite{Fox2013} performed similar study for 10 SNe IIn discovered
between 2005 and 2008, SN~2010jl, and seven other SNe.
Here, we compare the parameters of OGLE-2012-SN-006 with that of the SNe IIn
studied by \cite{Fox2011,Fox2013}, and some SNe
presented in other references.

\begin{figure}[tbph]
\vspace{80pt}
\begin{center}
\includegraphics[width=0.45\textwidth,angle=0]{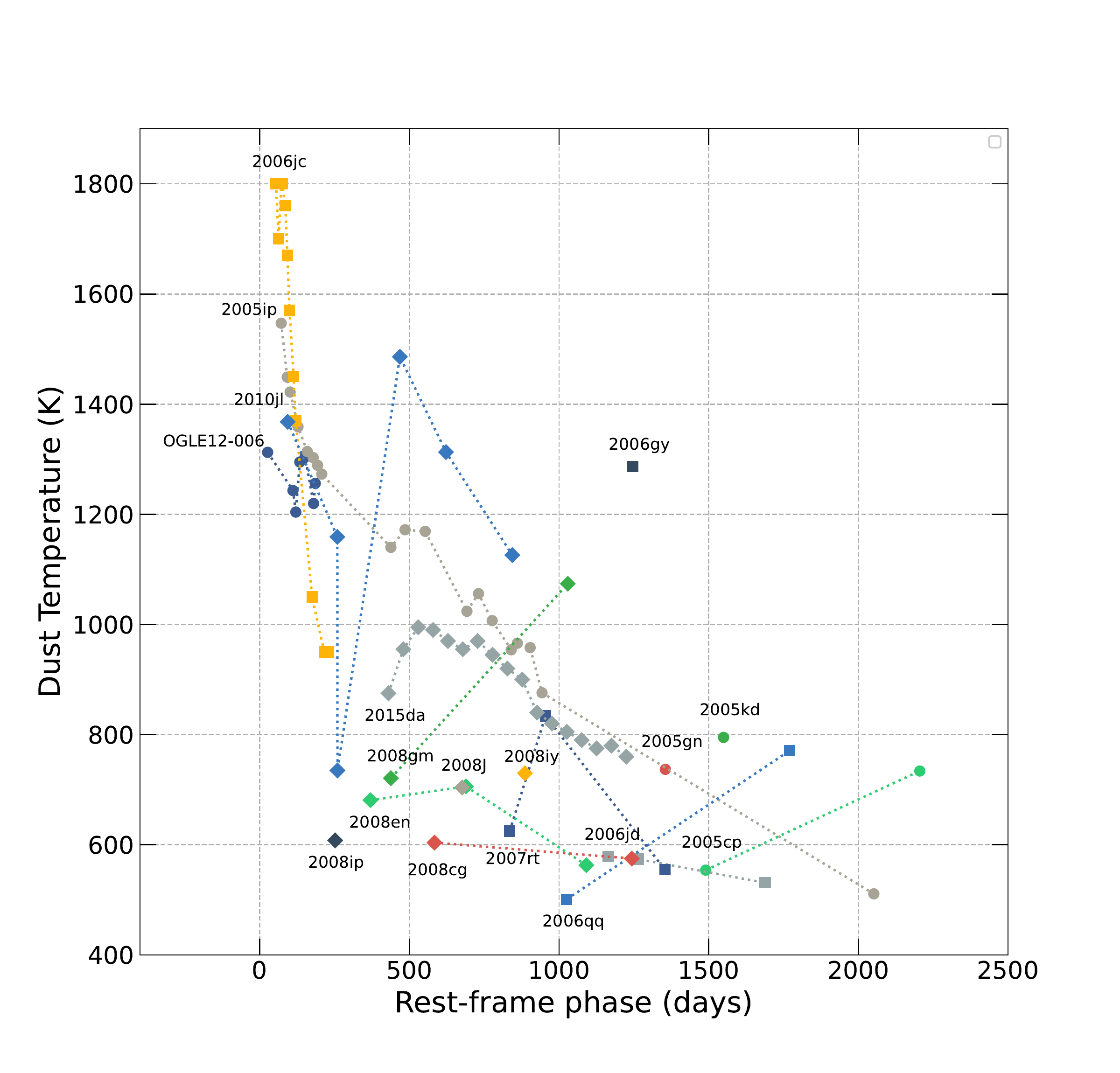}
\includegraphics[width=0.45\textwidth,angle=0]{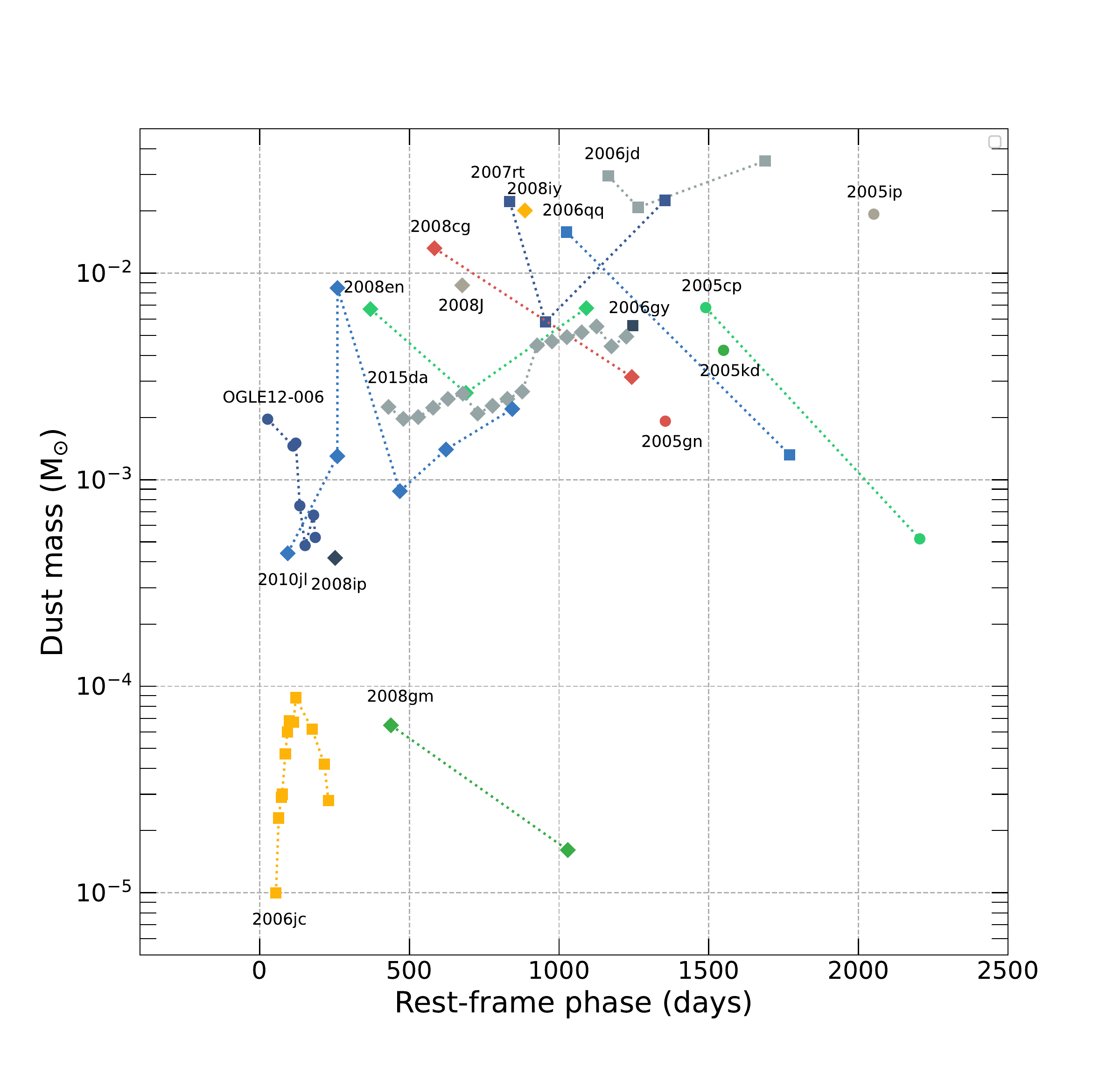}
\includegraphics[width=0.45\textwidth,angle=0]{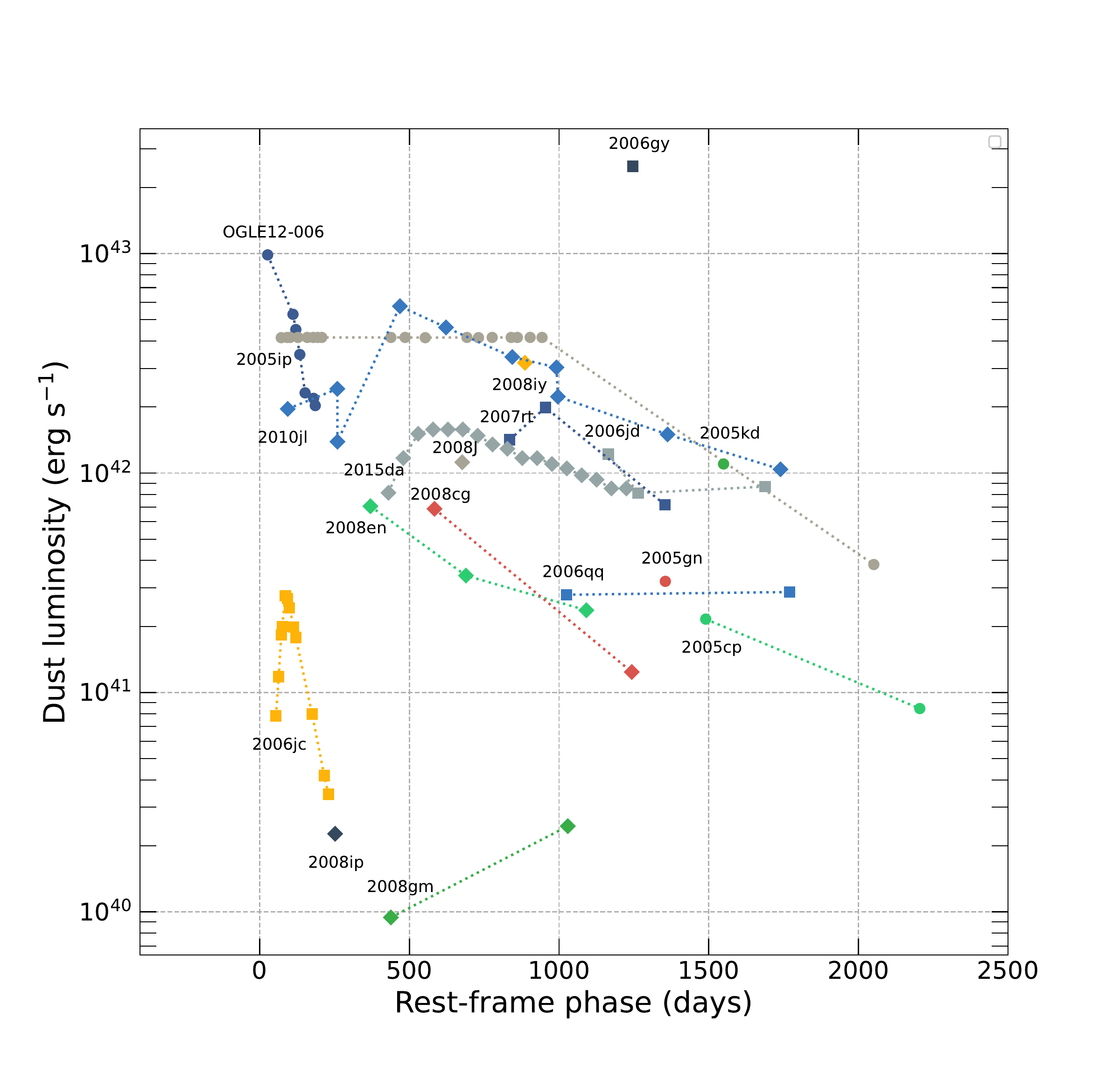}
\includegraphics[width=0.45\textwidth,angle=0]{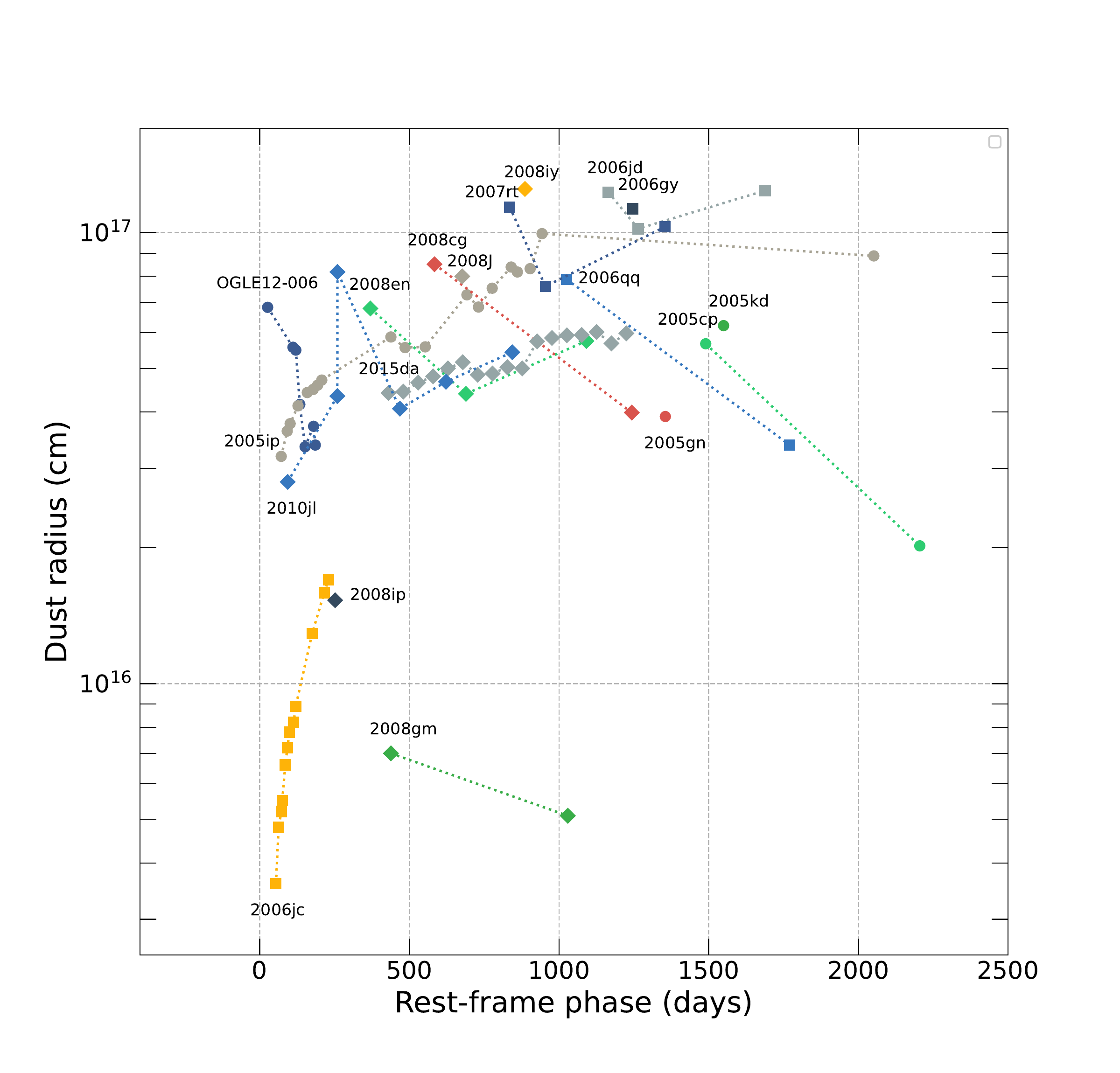}
\end{center}
\caption{The temperatures (the top-left panel),
the masses (the top-right panel),
the luminosities (the bottom-left panel), and
the radii (the bottom-right panel)
of the dust shells of OGLE-2012-SN-006,
SN~2006jc, and 16 SNe IIn.
The references can be found in the text.}
\label{fig:compar}
\end{figure}

In Figure \ref{fig:compar}, we plot the temperatures, radii, luminosities,
and masses of the dust component of OGLE-2012-SN-006 at all epochs
and these parameters of SN~2006jc \citep{Mattila2008} as well as 16 type IIn SNe
(SN~2005ip, \citealt{Fox2009,Fox2013}; SN~2005gn, SN~2008ip, SN~2008J, \citealt{Fox2011};
SN~2005cp, SN~2006jd, SN~2006qq, SN~2007rt, SN~2008cg, SN~2008en, SN~2008gm,
\citealt{Fox2011,Fox2013}; SN~2005kd, SN~2008iy, \citealt{Fox2013};
SN~2010jl, \citealt{Fox2013,Sarangi2018}; SN~2015da, \citealt{Tartaglia2020}; SN~2006gy, \citealt{Fox2013}).
By comparing the quantities of these SNe, one can get more interesting information.

The temperature of the dust of OGLE-2012-SN-006 is comparable to that of SNe~2006jc,
2005ip, 2010jl, and 2006gy at similar epochs, and higher than that of most of the SNe in
the figure. This might be due to the fact that the IR data of the latter
were obtained at late or very late epochs when the dust had cooled to lower temperatures.

The radius of the dust shell surrounding OGLE-2012-SN-006 is comparable to that of
most of SNe in the figure and larger than that of SNe~2006jc, 2008gm, and 2008ip.
As pointed out by \citet{Fox2011}, the MIR emission of the most SNe IIn
(SN~2005cp, SN~2005gn, SN~2006jd, SN~2006qq, SN~2007rt,
SN~2008cg, SN~2008en, SN~2008gm, SN~2008ip, and SN~2008J)
in their sample was produced by preexisting dust. The fact that the radius of the dust shell of
OGLE-2012-SN-006 is comparable to that of the SNe studied by \citet{Fox2011}
provides further support to the conclusion that the NIR excesses
of OGLE-2012-SN-006 might be produced by preexisting dust.

The luminosity of the dust of OGLE-2012-SN-006 is lower than that
of SN~2006gy, which is a superluminous SN IIn, and its peak is
more luminous than the luminosities of the dust of all other SNe in the figure.

The mass of the dust surrounding OGLE-2012-SN-006 is
$\sim$$0.5-2.0\times10^{-3} M_\odot$, about one magnitude
lower than that of almost all SNe in the figure, but
about one magnitude higher than that of SNe 2006jc and 2008gm.

\section{Conclusion}
\label{sec:Con}

It is widely believed that the IR excesses of SEDs of SNe might
be produced by the preexisting and/or newly formed dust shells,
so the early-time optical$-$NIR or late-time NIR/MIR SEDs
of SNe would provide important clues for diagnosing the dust formation
and make it possible to determine the physical properties
of the dust associated with the ejecta and/or the progenitors
of the SNe.

The aim of this paper is searching for the evidence of
dust formation of SNe Ibn.
To date, only six SNe Ibn have been observed in both optical and NIR bands.
Prior to this study, two SNe Ibn (SN~2006jc and PS1-12sk) having NIR photometry
have been researched. The optical$-$NIR SEDs of these two SNe Ibn exhibit different
NIR features: SN~2006jc shows early-time NIR excesses that have been interpreted by the
dust emission \citep{Mattila2008}, while the optical$-$NIR SEDs of PS1-12sk do not
have evident IR excesses since its SEDs can be matched by the
single-component blackbody model \citep{Sanders2013}.
In this paper, we investigate the optical$-$NIR SEDs of four
other Type Ibn SNe (SN~2010al, OGLE-2012-SN-006, LSQ13ddu, and SN~2015G)
having optical$-$NIR photometry.

By using the single-component blackbody model, we find that the SEDs of SN~2010al,
LSQ13ddu, and SN~2015G can be well explained by the model. The results indicate that
the evidence of the significant dust formation associated with
these three SNe is absent.
In contrast, the optical$-$NIR SEDs of OGLE-2012-SN-006 cannot be reproduced
by the single-component blackbody model and the NIR excesses of the SEDs of
OGLE-2012-SN-006 are obvious.

Therefore, we invoke the double-component model assuming that the SEDs were produced by
the SN photosphere and the heated dust to fit the SEDs of OGLE-2012-SN-006
at different epochs. To get the upper limits of the masses of the dust of SN~2010al,
LSQ13ddu, and SN~2015G, we also employ this model to fit their optical$-$NIR SEDs.

We find that the IR excesses of the SEDs of OGLE-2012-SN-006 can
be interpreted by the double-component model.
The best-fitting temperatures of the two classes of dust (graphite and silicate)
at different epochs favor the graphite dust, and the derived mass and temperature
of the graphite dust are $\sim$$0.5-2.0\times10^{-3}~M_\odot$ and $\sim$$1200-1300$ K,
respectively. This indicates that OGLE-2012-SN-006 is
the another SN Ibn showing evidence of the dust formation.
On the other hand, our fits for the SEDs of SN~2010al, LSQ13ddu, and SN~2015G
suggest that the upper limits of masses of possible dust of these three SNe Ibn are
$1.45\times 10^{-5}~M_\odot$, $5.9\times 10^{-7}~M_\odot$, and $2.4\times 10^{-7}~M_\odot$,
respectively.

A further analysis for the dust of OGLE-2012-SN-006 suggests that the radius of
the optically thin dust shell is $\sim$$6.8\times 10^{16}$cm which is significantly larger
than the radius of the SN ejecta at day 26.5 ($\sim$$1.9 \times 10^{15}$ cm),
indicating that the dust shell is preexisting one.


We compare the temperature, the radii, the luminosity,
as well as the mass of the dust
shell of OGLE-2012-SN-006 to that of SN~2006jc and 16 SNe IIn
(SN~2005cp, SN~2005gn, SN~2005ip, SN~2005kd,
SN~2006gy, SN~2006jd, SN~2006qq, SN~2007rt,
SN~2008cg, SN~2008en, SN~2008gm, SN~2008ip,
SN~2008iy, SN~2008J, SN~2010jl, and SN~2015da),
finding some additional information.
The temperature of the dust of OGLE-2012-SN-006 is comparable to that of SNe~2006jc,
2005ip, 2010jl, and 2006gy, and higher than that of most of the 17 SNe;
the radius of the dust shell of OGLE-2012-SN-006
is comparable to that of most SNe IIn and larger than that of SN~2006jc, SN~2008gm, and 2008ip;
the peak luminosity of the dust of OGLE-2012-SN-006
is higher than the luminosities of the dust of all other SNe, except for that of SN~2006gy;
the mass of the dust of OGLE-2012-SN-006 ($\sim$$10^{-3} M_\odot$) is about 1/10 times that of
most SNe IIn in the sample, but about 10 times that of SNe 2006jc and 2008gm.


According to the fits for the SEDs of two SNe Ibn (SN~2006jc and PS1-12sk) having optical$-$NIR
photometry coverage, one might infer that the percentage of SNe Ibn having IR excesses produced by
the heated dust is 50\%. However, the sample containing only two events was too small, making it
impossible to get a valid conclusion.
Our study, together with the previous studies, indicates that at least about 1/3
of SNe Ibn show evidence for dust formation.
\footnote{We caution that this fraction is approximate (because it is based on
a small sample size of six) and should be regarded as a lower limit (since
the evidence of the dust formation of many SNe was found by
analyzing the late-time observations, while all the SNe Ibn
we study lack late-time observations.)}

We caution, however, that the sample we collect is still rather small.
The future optical and IR observations for more SNe Ibn would
provide a larger sample, making it possible to perform a systematic study
for the dust formation of SNe Ibn and get a stronger conclusion
about the ratio of the SNe Ibn with significant dust formation to all SNe Ibn.

\acknowledgments
We thank the anonymous referee for helpful comments and
suggestions that have allowed us to improve this manuscript.
This work is supported by National Natural Science Foundation of China
(grants 11963001, 11533003, 11603006, 11673006, 11851304, 11973020 (C0035736), U1731239, and U1938201),
Guangxi Science Foundation (grants 2016GXNSFCB380005, 2016GXNSFFA380006,
and 2017GXNSFFA198008, AD17129006, 2018GXNSFGA281007, 2019JJD110006, and 2019AC20334), and the Bagui Young
Scholars Program (LHJ).

\clearpage

\vspace{30pt}
\setlength{\tabcolsep}{2pt}
\begin{table*}
\begin{center}
\caption{The information on the SNe in the sample}
\label{table:details}
\begin{tabular}{cccccccccccc}
\hline
\hline
Name & R.A.  & Decl. & Redshift &  Discovery Date & Obs. Filters & References$^a$ \\
  & (J2000)  &  (J2000)   &   &    &   &   \\
\hline
\hline
SN~2010al   & \ra{08}{14}{15}{91}& \dec{+18}{26}{18}{2}  & 0.017 & Mar 13.03 2010 & UVOT filters$^b$, $U$, $B$, $V$, $R$, $I$, $J$, $H$, $K_s$  & 1 \\
OGLE-2012-SN-006  & \ra{03}{33}{34}{79}& \dec{-74}{23}{40}{1}  & 0.06  & Oct 7.34 2012  & $U$, $B$, $V$, $R$, $I$, $J$, $H$, $K_s$                & 2 \\
LSQ13ddu    & \ra{03}{58}{49}{09}& \dec{+29}{25}{11}{8}  & 0.058 & Nov 26.3 2013  & UVOT filters$^b$, $U$, $B$, $V$, $g$, $r$, $i$, $z$, $Y$, $J$, $H$, $K_s$ & 3  \\
SN~2015G    & \ra{20}{37}{25}{58}& \dec{+66}{07}{11}{5}  & 0.005 & Mar 23.78 2015 & $B$, $V$, $R$, $I$, $J$, $H$, $K_s$                 & 4 \\
\hline
\hline
\end{tabular}
\end{center}
$^a$ References: (1) \cite{Pas2015a}; (2) \cite{Pas2015b}; (3) \cite{Clark2020}; (4) \cite{Shiv2017}. \\
$^b$ $uvw2$, $uvm2$, $uvw1$, $u$, $b$, and $v$.
\end{table*}

\clearpage


\begin{table*}
\footnotesize
\setlength{\tabcolsep}{1.8pt}
\renewcommand{\arraystretch}{1.05}

\begin{center}
\caption{\label{table:SED_PARAM-double}The best-fitting parameters of the Blackbody plus Graphite (Silicate) model for the SEDs of the four SNe Ibn at the different rest-frame epochs. }

\begin{tabular}{c c c c c c c c c c c c c}

\hline\hline\noalign{\smallskip}

&\multicolumn{5}{c}{\textbf{Blackbody plus Graphite}} & \multicolumn{5}{c}{\textbf{Blackbody plus Silicate}}\\\noalign{\smallskip}
\cmidrule(lr){2-6} \cmidrule(lr){7-11}

\colhead{Phase\footnote{All the phases are relative to the possible corresponding explosion days that are supposed to be respectively JD 2455268 (SN~2010al), JD 2456203.8 (OGLE-2012-SN-006), JD 2457080.5 (SN~2015G), and JD 2456621.4 (LSQ13ddu).}} & \colhead{$T_{\rm ph}$}&\colhead{$R_{\rm ph}$}&\colhead{$T_{\rm d}$}&\colhead{$M_{\rm d}$}&\colhead{$\chi^{\rm 2}$/dof}&\colhead{$T_{\rm ph}$}&\colhead{$R_{\rm ph}$}&\colhead{$T_{\rm d}$} & \colhead{$M_{\rm d}$}  & \colhead{$\chi^{\rm 2}$/dof} \\
& (K) & (10$^{15}$ cm) &  (K) & (10$^{-6}$M$_{\odot}$) &  & (K) & (10$^{15}$ cm) & (K) & (10$^{-6}$M$_{\odot}$) & \\\noalign{\smallskip} \hline \noalign{\smallskip}

\multicolumn{11}{c}{\textbf{SN 2010al}}\\\noalign{\smallskip} \hline \noalign{\smallskip}

11.3 d & $13666.97^{+625.6}_{-520.7}$ & $0.9^{+0.1}_{-0.1}$ & $521.75^{+1251.8}_{-351.9}$ & $1.79^{+854.8}_{-1.8}$ & 2.88 & $13648.11^{+599.3}_{-512.0}$ & $0.9^{+0.1}_{-0.1}$ & $555.96^{+1270.2}_{-380.0}$ & $2.79^{+1250.4}_{-2.8}$ & 2.87\\\noalign{\smallskip} \hline \noalign{\smallskip}
12.3 d & $10504.2^{+1319.3}_{-1119.4}$ & $1.25^{+0.3}_{-0.2}$ & $1651.93^{+484.4}_{-869.1}$ & $14.53^{+67.9}_{-10.6}$ & 0.76 & $10576.04^{+1619.4}_{-1214.4}$ & $1.23^{+0.3}_{-0.2}$ & $2010.38^{+693.6}_{-1100.0}$ & $83.62^{+272.4}_{-55.2}$ & 0.96\\\noalign{\smallskip} \hline \noalign{\smallskip}
14.3 d & $10025.75^{+531.8}_{-459.4}$ & $1.35^{+0.1}_{-0.1}$ & $639.69^{+1099.0}_{-451.8}$ & $4.27^{+1021.6}_{-4.3}$ & 2.15 & $10042.15^{+522.3}_{-455.8}$ & $1.35^{+0.1}_{-0.1}$ & $714.51^{+1224.4}_{-524.0}$ & $16.06^{+1830.1}_{-16.1}$ & 2.11\\\noalign{\smallskip} \hline \noalign{\smallskip}
16.2 d & $9730.24^{+193.0}_{-188.6}$ & $1.41^{+0.1}_{-0.1}$ & $452.65^{+1139.4}_{-289.7}$ & $0.72^{+497.3}_{-0.7}$ & 0.69 & $9731.01^{+184.9}_{-178.4}$ & $1.41^{+0.1}_{-0.1}$ & $479.14^{+1097.3}_{-314.4}$ & $1.06^{+710.4}_{-1.1}$ & 0.68\\\noalign{\smallskip} \hline \noalign{\smallskip}
23.0 d & $7667.52^{+119.4}_{-122.4}$ & $1.82^{+0.1}_{-0.1}$ & $394.65^{+835.8}_{-240.7}$ & $0.41^{+680.2}_{-0.4}$ & 2.08 & $7667.84^{+121.4}_{-122.3}$ & $1.83^{+0.1}_{-0.1}$ & $449.84^{+967.8}_{-287.1}$ & $0.77^{+786.4}_{-0.8}$ & 2.08\\\noalign{\smallskip} \hline \noalign{\smallskip}
26.0 d & $7263.98^{+119.3}_{-126.0}$ & $1.8^{+0.1}_{-0.1}$ & $439.33^{+826.7}_{-278.2}$ & $0.85^{+1165.2}_{-0.9}$ & 2.87 & $7265.26^{+119.9}_{-118.7}$ & $1.8^{+0.1}_{-0.1}$ & $474.91^{+893.0}_{-311.3}$ & $1.25^{+1500.4}_{-1.2}$ & 2.87\\\noalign{\smallskip} \hline \noalign{\smallskip}
29.8 d & $6786.31^{+129.9}_{-130.3}$ & $1.73^{+0.1}_{-0.1}$ & $378.14^{+822.4}_{-226.6}$ & $0.31^{+557.1}_{-0.3}$ & 1.66 & $6786.59^{+130.1}_{-130.5}$ & $1.73^{+0.1}_{-0.1}$ & $424.24^{+921.3}_{-264.6}$ & $0.64^{+756.0}_{-0.6}$ & 1.66\\\noalign{\smallskip} \hline \noalign{\smallskip}
37.0 d & $6467.84^{+153.6}_{-152.3}$ & $1.21^{+0.1}_{-0.1}$ & $369.78^{+812.5}_{-217.6}$ & $0.24^{+558.2}_{-0.2}$ & 1.49 & $6463.44^{+151.6}_{-151.5}$ & $1.21^{+0.1}_{-0.1}$ & $404.2^{+917.2}_{-246.7}$ & $0.49^{+678.2}_{-0.5}$ & 1.49\\\noalign{\smallskip} \hline \noalign{\smallskip}
40.7 d & $6215.12^{+249.2}_{-241.7}$ & $1.05^{+0.1}_{-0.1}$ & $384.93^{+883.8}_{-231.2}$ & $0.36^{+563.9}_{-0.4}$ & 0.36 & $6216.45^{+242.3}_{-243.8}$ & $1.05^{+0.1}_{-0.1}$ & $431.94^{+1006.5}_{-272.4}$ & $0.58^{+667.5}_{-0.6}$ & 0.36\\\noalign{\smallskip} \hline \noalign{\smallskip}
43.6 d & $6193.35^{+363.7}_{-377.9}$ & $0.9^{+0.1}_{-0.1}$ & $404.42^{+903.8}_{-248.6}$ & $0.43^{+765.8}_{-0.4}$ & 0.64 & $6201.26^{+356.7}_{-379.5}$ & $0.9^{+0.1}_{-0.1}$ & $446.64^{+998.5}_{-284.3}$ & $0.93^{+992.6}_{-0.9}$ & 0.64\\\noalign{\smallskip} \hline \noalign{\smallskip}
59.5 d & $6069.16^{+218.2}_{-202.6}$ & $0.37^{+0.0}_{-0.0}$ & $1141.56^{+216.4}_{-254.8}$ & $7.88^{+33.2}_{-5.1}$ & 3.04 & $6070.44^{+224.0}_{-204.0}$ & $0.37^{+0.0}_{-0.0}$ & $1329.58^{+292.0}_{-318.6}$ & $40.51^{+150.6}_{-25.8}$ & 3.12\\\noalign{\smallskip} \hline \noalign{\smallskip}

\multicolumn{11}{c}{\textbf{OGLE-2012-SN-006}}\\\noalign{\smallskip} \hline \noalign{\smallskip}
26.5 d & $7783.53^{+475.4}_{-419.3}$ & $1.92^{+0.2}_{-0.2}$ & $1312.57^{+47.3}_{-47.6}$ & $1963.09^{+593.9}_{-441.8}$ & 1.62 & $7986.53^{+541.6}_{-524.4}$ & $1.83^{+0.2}_{-0.2}$ & $1505.2^{+61.2}_{-66.7}$ & $12810.53^{+3613.9}_{-3037.9}$ & 1.91\\\noalign{\smallskip} \hline \noalign{\smallskip}
111.1 d & $6504.53^{+306.2}_{-290.3}$ & $1.6^{+0.1}_{-0.1}$ & $1243.17^{+42.7}_{-41.0}$ & $1456.3^{+406.0}_{-321.1}$ & 3.74 & $6688.26^{+364.1}_{-347.4}$ & $1.49^{+0.2}_{-0.1}$ & $1456.82^{+59.2}_{-58.3}$ & $7987.22^{+2188.0}_{-1810.3}$ & 2.61\\\noalign{\smallskip} \hline \noalign{\smallskip}
120.6 d & $6541.24^{+263.3}_{-246.1}$ & $1.55^{+0.1}_{-0.1}$ & $1204.05^{+57.2}_{-54.9}$ & $1502.76^{+599.3}_{-428.7}$ & 3.57 & $6659.75^{+300.4}_{-271.1}$ & $1.49^{+0.1}_{-0.1}$ & $1417.71^{+78.9}_{-74.5}$ & $7821.45^{+3119.6}_{-2207.8}$ & 2.79\\\noalign{\smallskip} \hline \noalign{\smallskip}
133.9 d & $7144.95^{+253.9}_{-237.9}$ & $1.28^{+0.1}_{-0.1}$ & $1294.96^{+66.0}_{-60.2}$ & $747.98^{+319.2}_{-229.8}$ & 3.9 & $7369.76^{+346.9}_{-293.0}$ & $1.2^{+0.1}_{-0.1}$ & $1515.52^{+86.5}_{-80.2}$ & $4346.93^{+1768.7}_{-1266.1}$ & 3.55\\\noalign{\smallskip} \hline \noalign{\smallskip}
151.6 d & $7974.58^{+431.7}_{-391.2}$ & $0.95^{+0.1}_{-0.1}$ & $1303.87^{+68.7}_{-66.5}$ & $479.87^{+214.3}_{-145.3}$ & 17.29 & $8116.14^{+471.8}_{-420.5}$ & $0.92^{+0.1}_{-0.1}$ & $1520.6^{+86.6}_{-82.0}$ & $2889.66^{+1165.1}_{-813.6}$ & 15.6\\\noalign{\smallskip} \hline \noalign{\smallskip}
179.9 d & $7208.97^{+430.4}_{-392.0}$ & $0.91^{+0.1}_{-0.1}$ & $1219.67^{+72.6}_{-69.4}$ & $673.45^{+357.6}_{-228.9}$ & 7.32 & $7346.22^{+482.6}_{-411.5}$ & $0.88^{+0.1}_{-0.1}$ & $1436.77^{+95.1}_{-91.3}$ & $3551.44^{+1777.3}_{-1140.6}$ & 6.24\\\noalign{\smallskip} \hline \noalign{\smallskip}
185.6 d & $6428.12^{+418.3}_{-391.8}$ & $1.07^{+0.1}_{-0.1}$ & $1256.21^{+53.6}_{-52.1}$ & $525.17^{+189.1}_{-135.0}$ & 9.91 & $6609.77^{+465.1}_{-462.5}$ & $1.0^{+0.2}_{-0.1}$ & $1477.91^{+74.5}_{-74.7}$ & $2828.39^{+993.5}_{-761.4}$ & 7.95\\\noalign{\smallskip} \hline \noalign{\smallskip}

\multicolumn{11}{c}{\textbf{LSQ13ddu}}\\\noalign{\smallskip} \hline \noalign{\smallskip}
3.1 d & $50540.99^{+6570.7}_{-5253.3}$ & $0.42^{+0.0}_{-0.0}$ & $2262.37^{+62.8}_{-62.9}$ & $10.54^{+1.5}_{-1.4}$ & 55.74 & $55602.67^{+10430.9}_{-7489.6}$ & $0.39^{+0.0}_{-0.0}$ & $2599.33^{+86.4}_{-90.3}$ & $111.44^{+15.8}_{-13.9}$ & 67.72\\\noalign{\smallskip} \hline \noalign{\smallskip}
14.5 d & $8067.59^{+276.9}_{-191.7}$ & $1.56^{+0.1}_{-0.1}$ & $449.84^{+1941.9}_{-288.3}$ & $0.59^{+257.7}_{-0.6}$ & 3.93 & $8037.6^{+189.7}_{-176.9}$ & $1.57^{+0.1}_{-0.1}$ & $414.05^{+990.7}_{-255.6}$ & $0.53^{+611.5}_{-0.5}$ & 3.92\\\noalign{\smallskip} \hline \noalign{\smallskip}

\multicolumn{11}{c}{\textbf{SN 2015G}}\\\noalign{\smallskip} \hline \noalign{\smallskip}
42.3 d & $6837.98^{+214.6}_{-212.7}$ & $0.44^{+0.0}_{-0.0}$ & $302.12^{+561.0}_{-159.3}$ & $0.14^{+482.0}_{-0.1}$ & 13.96 & $6836.47^{+216.0}_{-213.2}$ & $0.44^{+0.0}_{-0.0}$ & $331.73^{+634.4}_{-184.9}$ & $0.24^{+504.1}_{-0.2}$ & 13.96\\\noalign{\smallskip} \hline \noalign{\smallskip}
59.2 d & $6317.11^{+228.1}_{-230.1}$ & $0.36^{+0.0}_{-0.0}$ & $305.74^{+660.9}_{-162.5}$ & $0.11^{+415.1}_{-0.1}$ & 7.11 & $6312.44^{+229.9}_{-225.6}$ & $0.36^{+0.0}_{-0.0}$ & $327.27^{+645.9}_{-180.9}$ & $0.21^{+507.0}_{-0.2}$ & 7.1\\\noalign{\smallskip} \hline \noalign{\smallskip}
\end{tabular}
\end{center}
\quad\textbf{Note.} Here, $T_{\rm ph}$ is the temperature of the SN photosphere, $R_{\rm ph}$ is the radius of the SN photosphere, $T_{\rm d}$ is the temperature of the dust shell, and $M_{\rm d}$ is the mass of the dust shell.

\end{table*}


\clearpage
\begin{table*}

\vspace{30pt}
\centering
\tabletypesize{\scriptsize}

\caption{\label{table:SED_L}The derived values of the SN photosphere luminosities ($L_{\rm ph}$), the dust
luminosities ($L_{\rm d}$), and the lower limits of the radii ($R_{\rm d}$) of the dust shell surrounding
OGLE-2012-SN-006 at different epochs.}

\vspace{20pt}
\begin{tabular}{c c c c c c c}
\hline\hline\noalign{\smallskip}

\colhead{Phase} & \colhead{$L_{\rm ph}$} & \colhead{$L_{\rm d}$} & \colhead{$R_{\rm d}$} \\
(days)& ($\rm 10^{42}\ erg\ s^{-1}$)  & ($\rm 10^{42}\ erg\ s^{-1}$) & ($\rm 10^{16}\ cm$) \\\noalign{\smallskip}\hline\noalign{\smallskip}
26.5 & $9.65^{+0.5}_{-0.4}$ & $9.87^{+0.6}_{-0.6}$ & $>6.83^{+0.7}_{-0.6}$\\\noalign{\smallskip}\hline\noalign{\smallskip}
111.1 & $3.25^{+0.1}_{-0.1}$ & $5.29^{+0.3}_{-0.3}$ & $>5.57^{+0.5}_{-0.5}$\\\noalign{\smallskip}\hline\noalign{\smallskip}
120.6 & $3.16^{+0.1}_{-0.1}$ & $4.51^{+0.4}_{-0.4}$ & $>5.49^{+0.7}_{-0.6}$\\\noalign{\smallskip}\hline\noalign{\smallskip}
133.9 & $3.03^{+0.1}_{-0.1}$ & $3.47^{+0.4}_{-0.3}$ & $>4.16^{+0.6}_{-0.5}$\\\noalign{\smallskip}\hline\noalign{\smallskip}
151.6 & $2.6^{+0.1}_{-0.1}$ & $2.32^{+0.2}_{-0.2}$ & $>3.35^{+0.5}_{-0.4}$\\\noalign{\smallskip}\hline\noalign{\smallskip}
179.9 & $1.62^{+0.1}_{-0.0}$ & $2.19^{+0.2}_{-0.2}$ & $>3.72^{+0.6}_{-0.5}$\\\noalign{\smallskip}\hline\noalign{\smallskip}
185.6 & $1.39^{+0.0}_{-0.0}$ & $2.03^{+0.1}_{-0.1}$ & $>3.38^{+0.4}_{-0.3}$\\\noalign{\smallskip}\hline\noalign{\smallskip}
\end{tabular}
\end{table*}

\clearpage

\clearpage

\clearpage

\clearpage

\clearpage

\appendix
\setcounter{table}{0}
\setcounter{figure}{0}
\setcounter{equation}{0}
\renewcommand{\thetable}{A\arabic{table}}
\renewcommand{\thefigure}{A\arabic{figure}}
\renewcommand\theequation{A.\arabic{equation}}

Figures \ref{fig:2010corner}, \ref{fig:2012corner}, \ref{fig:2013corner}, and \ref{fig:2015corner} show the corner plots of the
models in the main text.

\clearpage

\begin{figure}
\vspace{80pt}
\centering
\includegraphics[width=0.48\textwidth,angle=0]{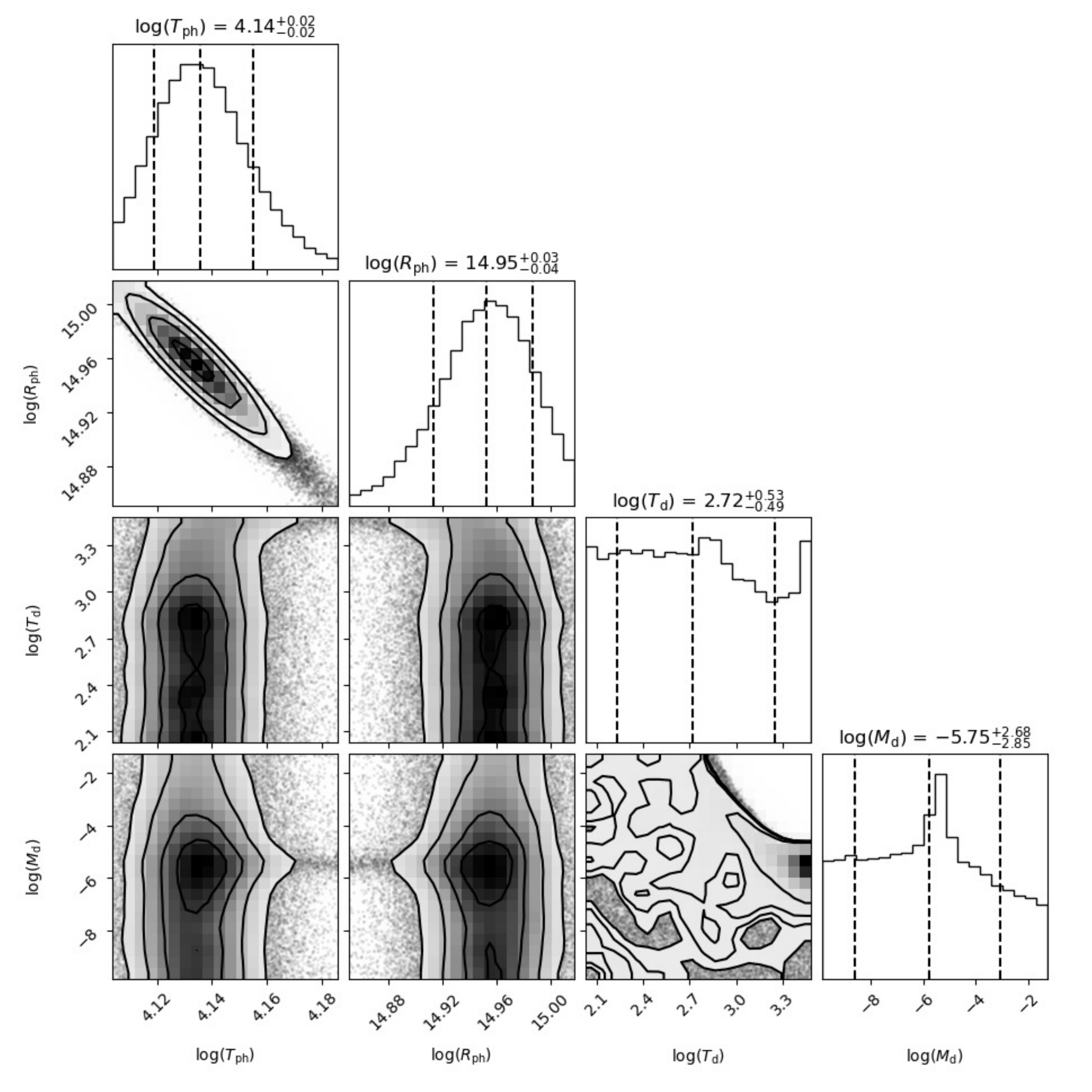}
\includegraphics[width=0.48\textwidth,angle=0]{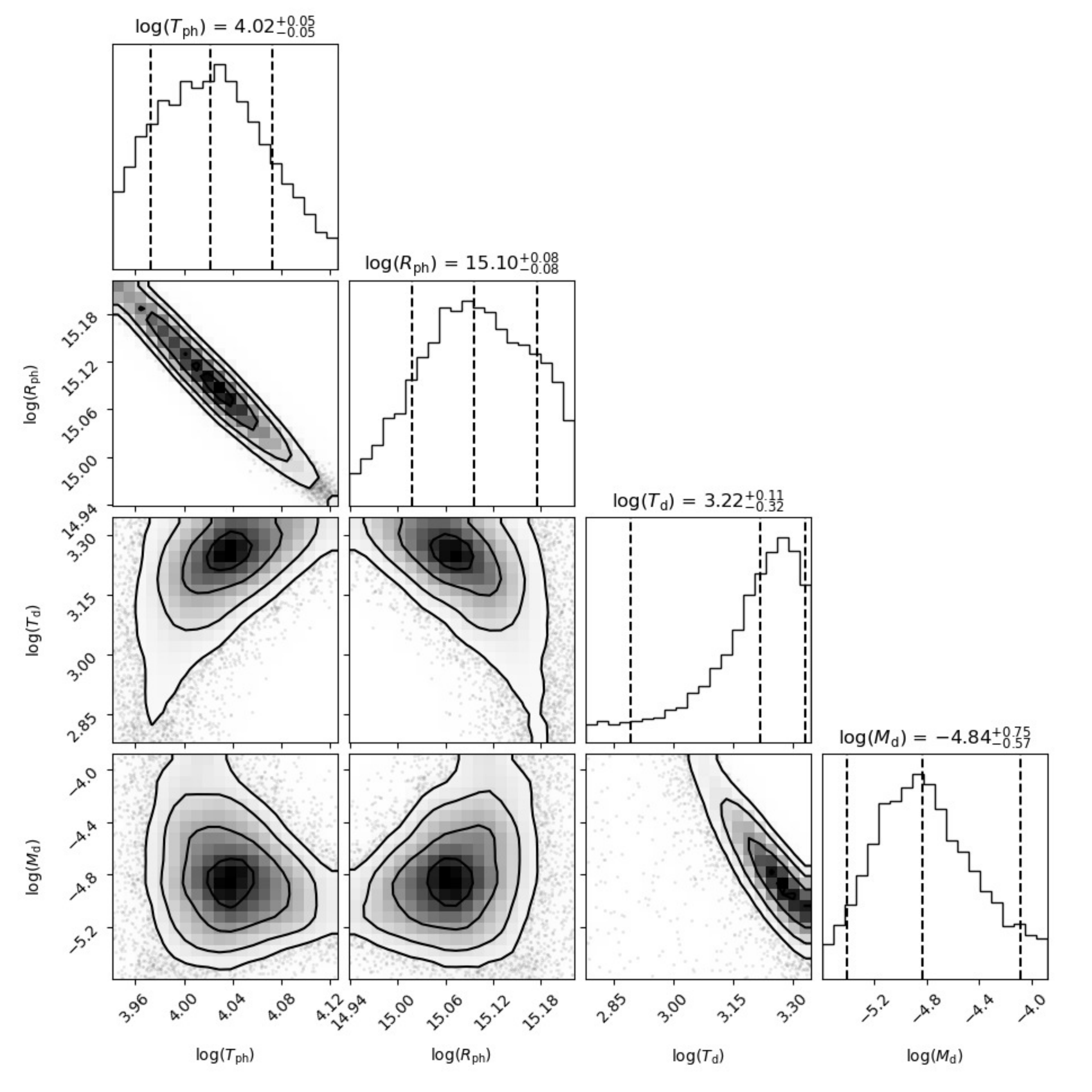}\\
\includegraphics[width=0.48\textwidth,angle=0]{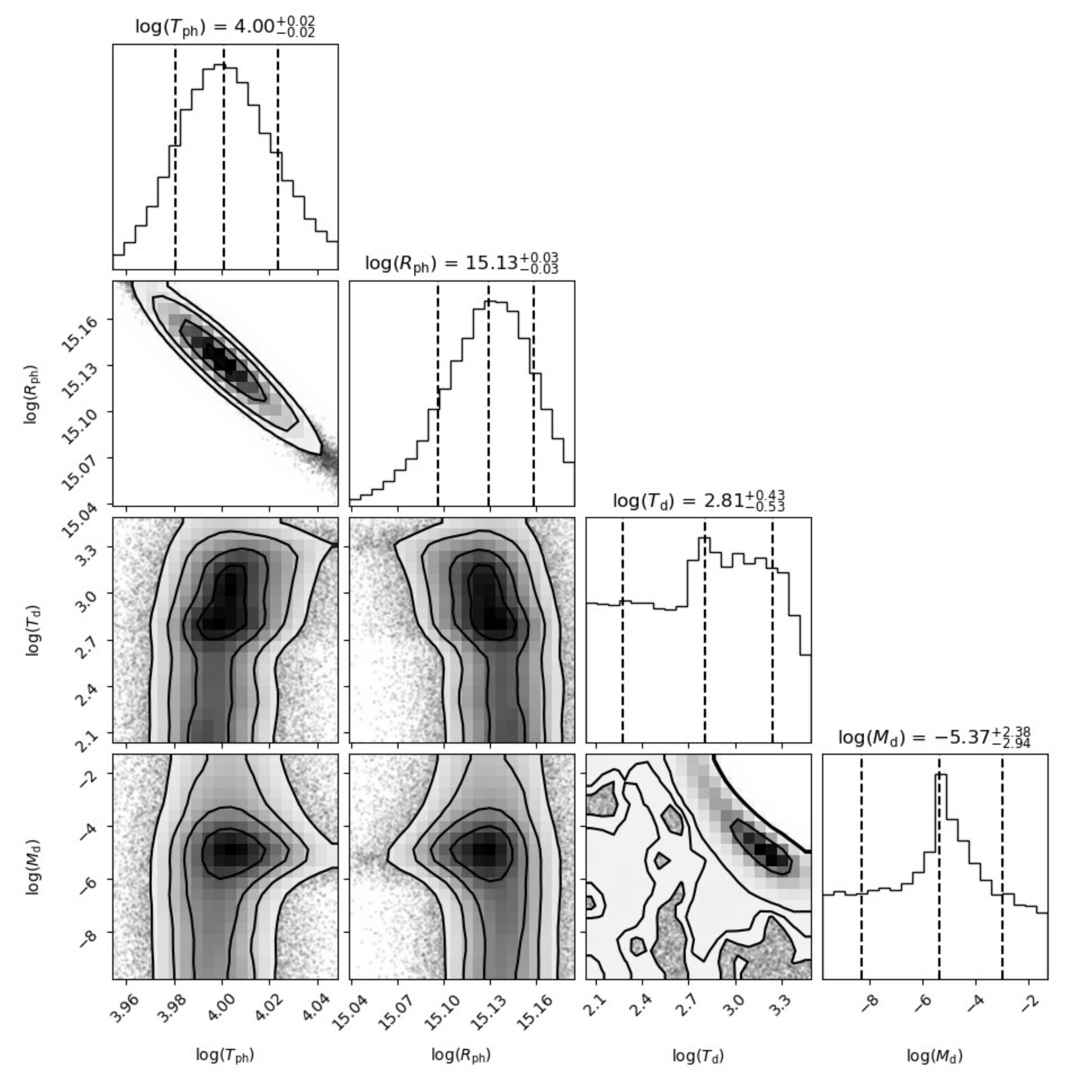}
\includegraphics[width=0.48\textwidth,angle=0]{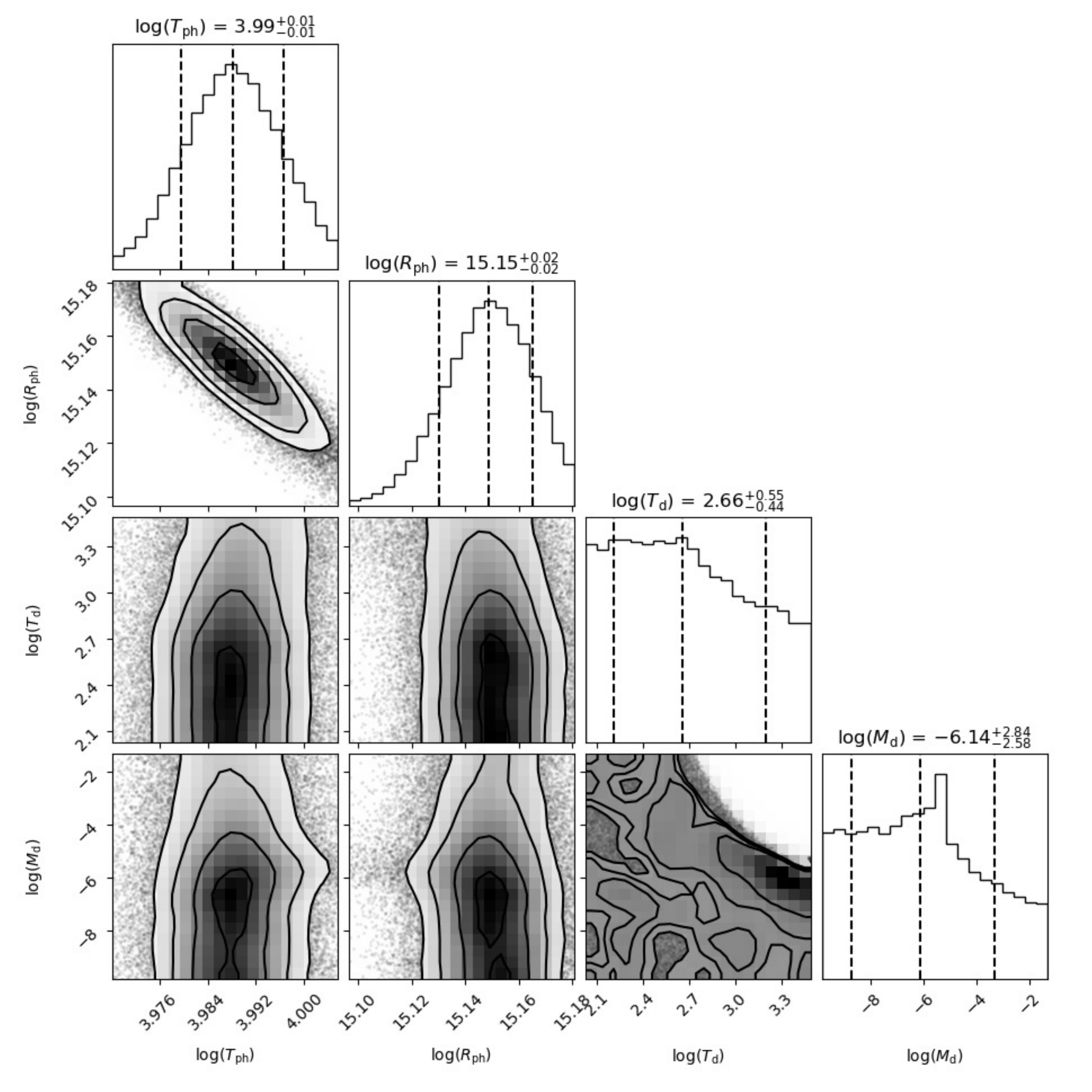}
\caption{The corner plots of the two-component model for SEDs of SN~2010al.}
\label{fig:2010corner}
\end{figure}

\begin{figure}
\ContinuedFloat
\vspace{80pt}
\centering
\includegraphics[width=0.48\textwidth,angle=0]{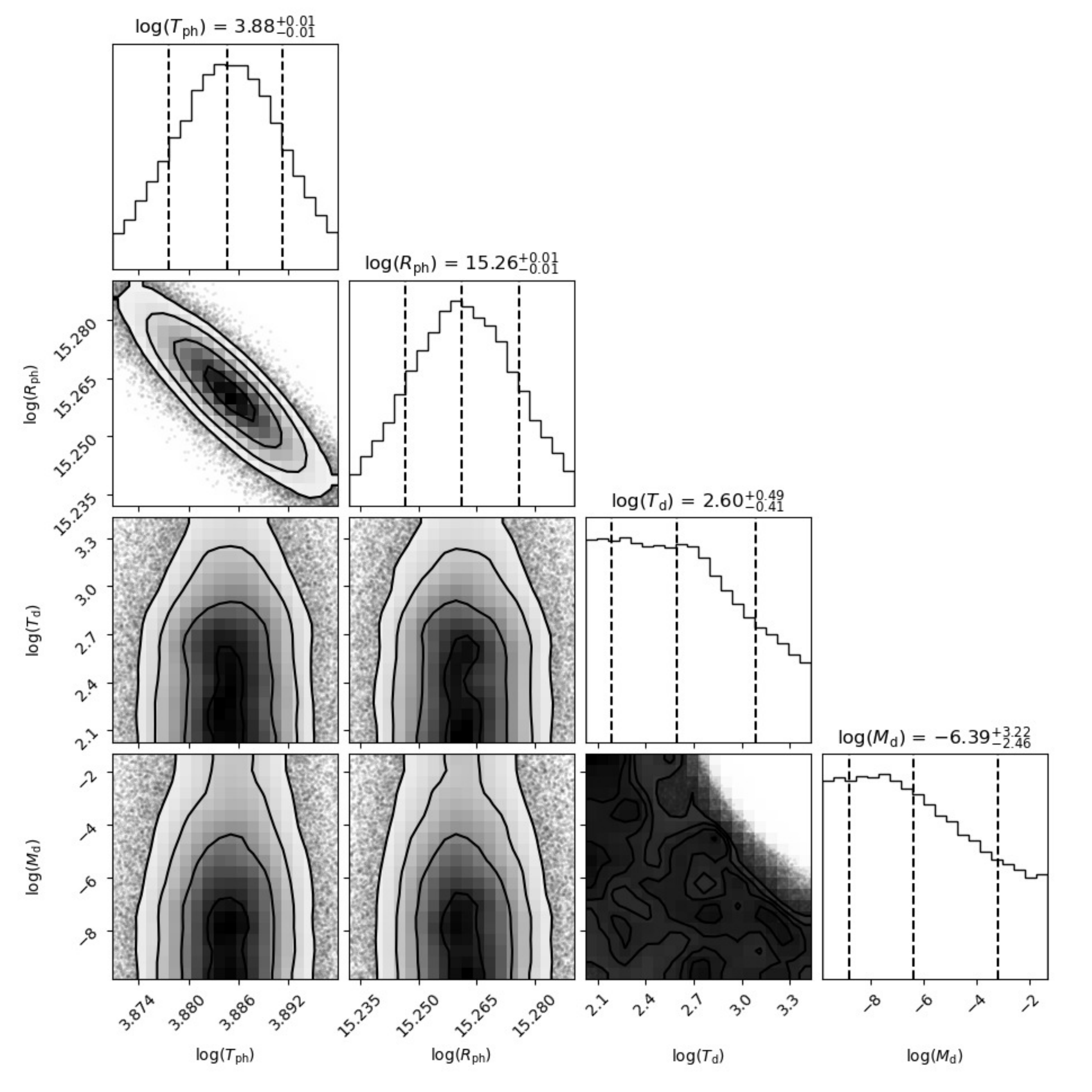}
\includegraphics[width=0.48\textwidth,angle=0]{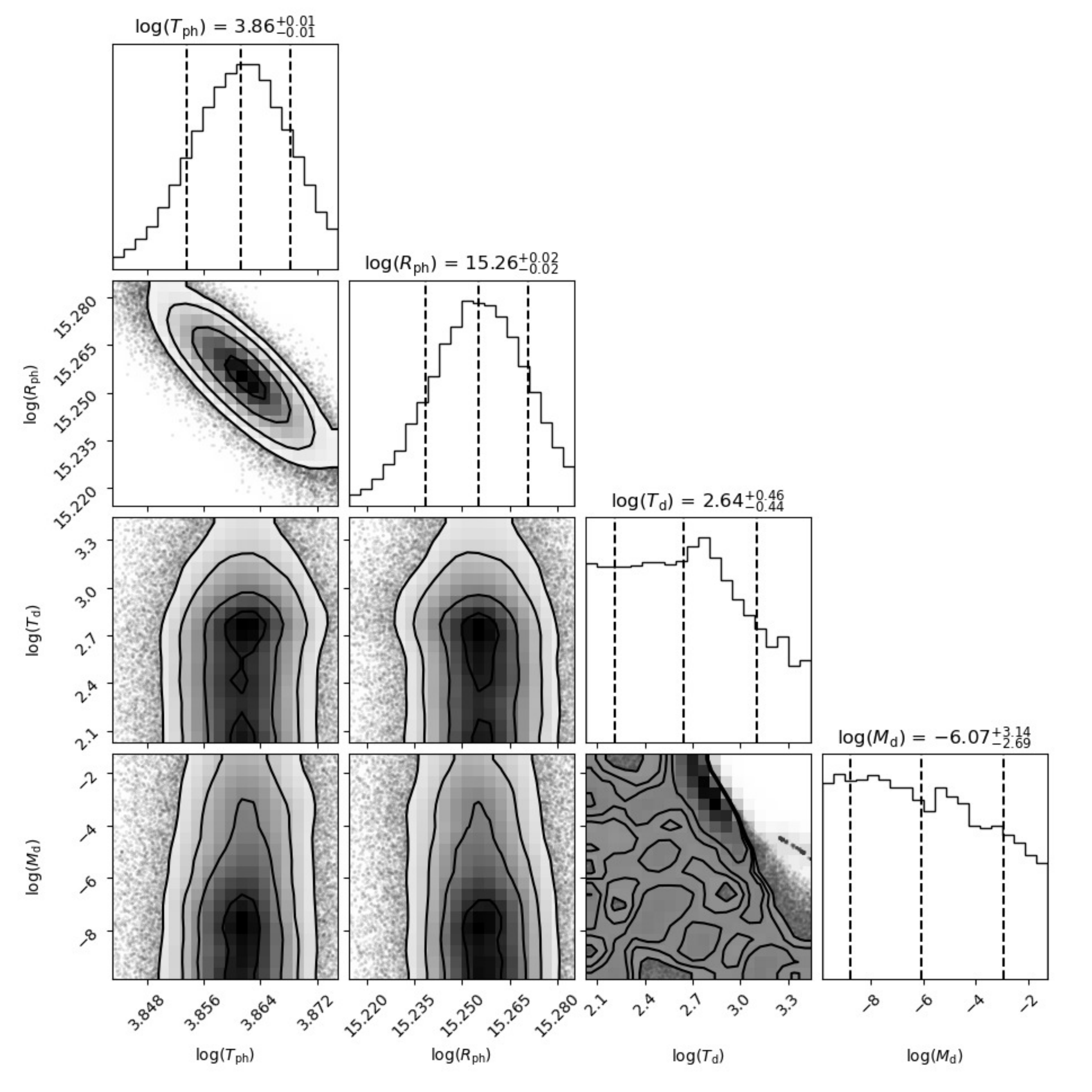}\\
\includegraphics[width=0.48\textwidth,angle=0]{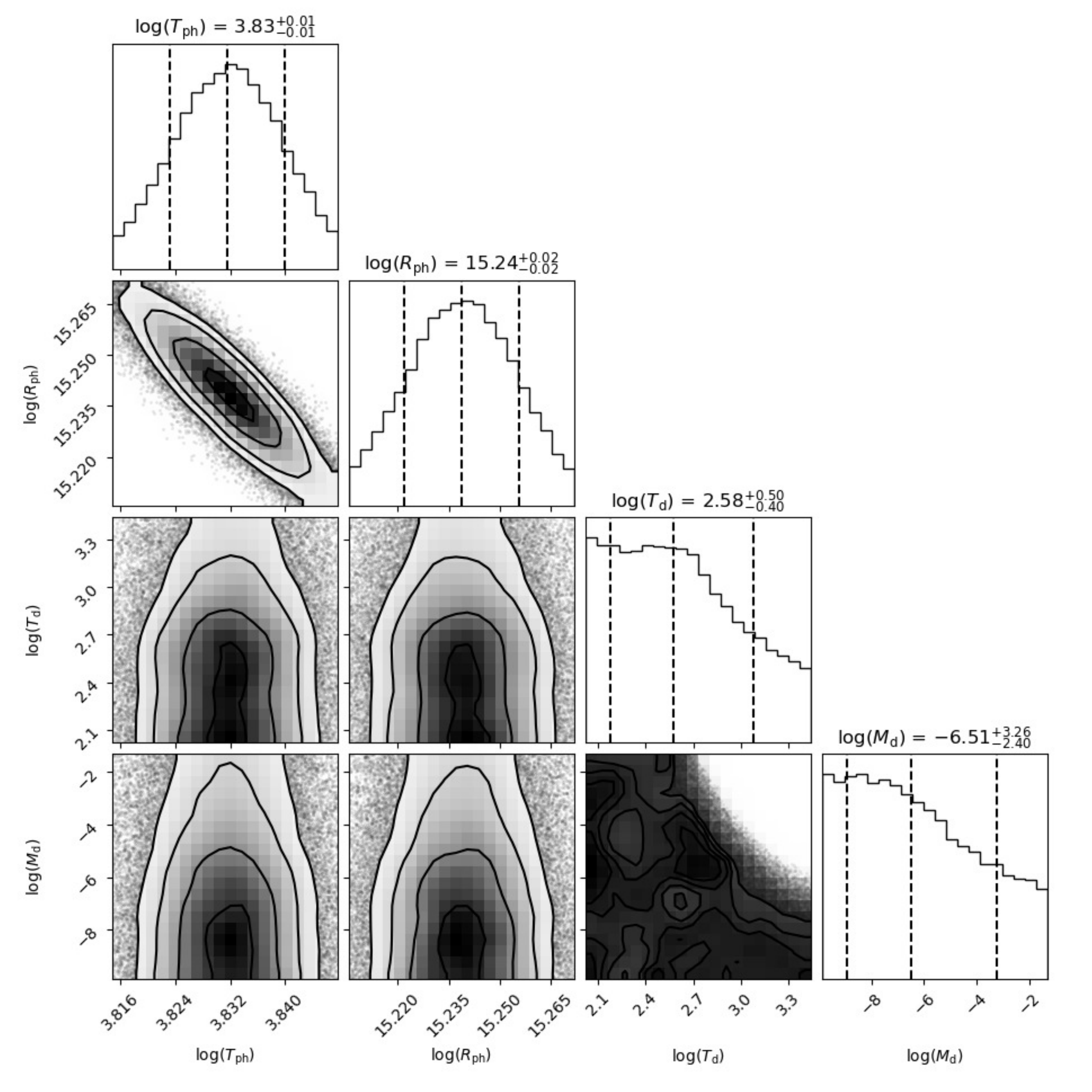}
\includegraphics[width=0.48\textwidth,angle=0]{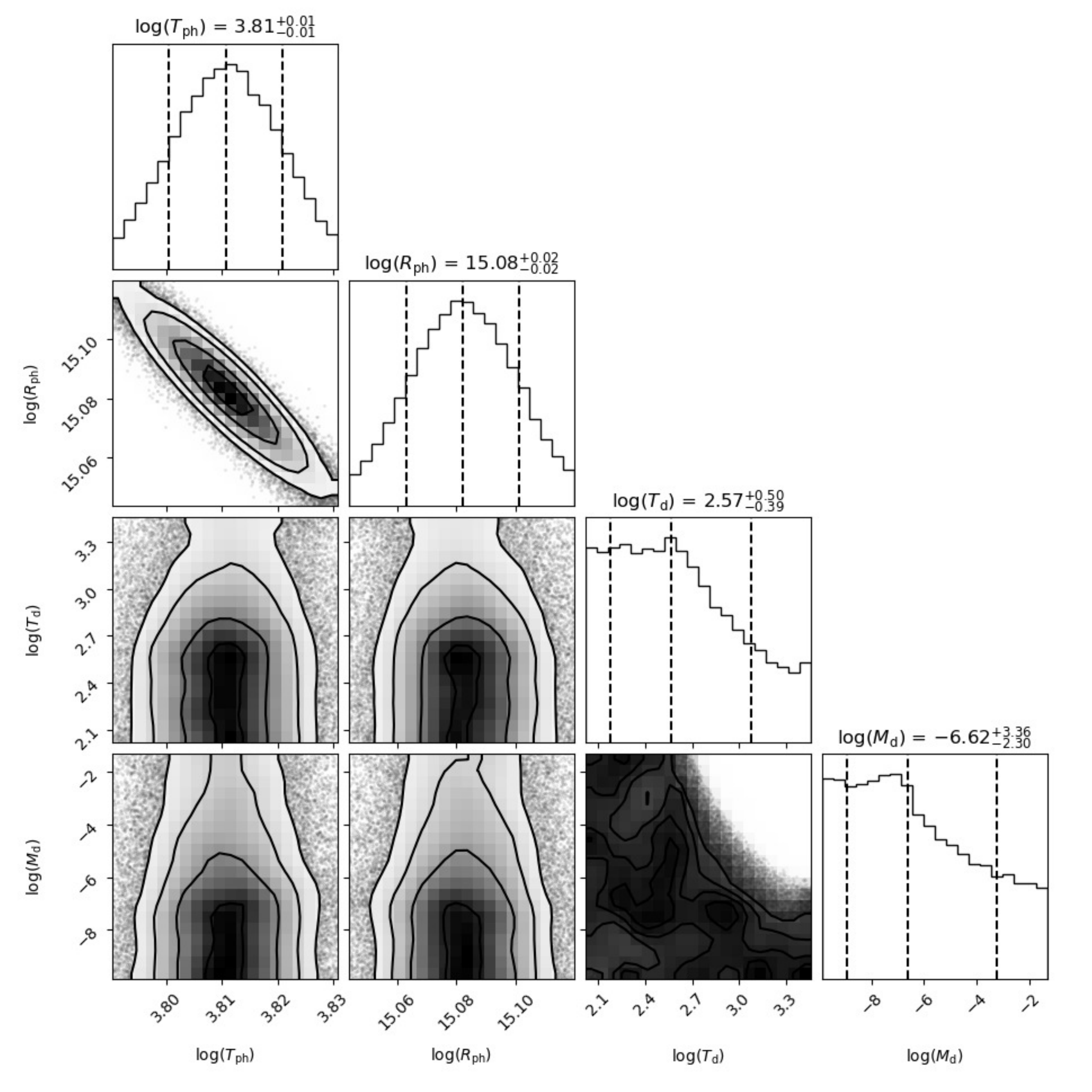}
\caption{The corner plots of the two-component model for SEDs of SN~2010al---continued.}
\end{figure}

\begin{figure}
\ContinuedFloat
\vspace{80pt}
\centering
\includegraphics[width=0.48\textwidth,angle=0]{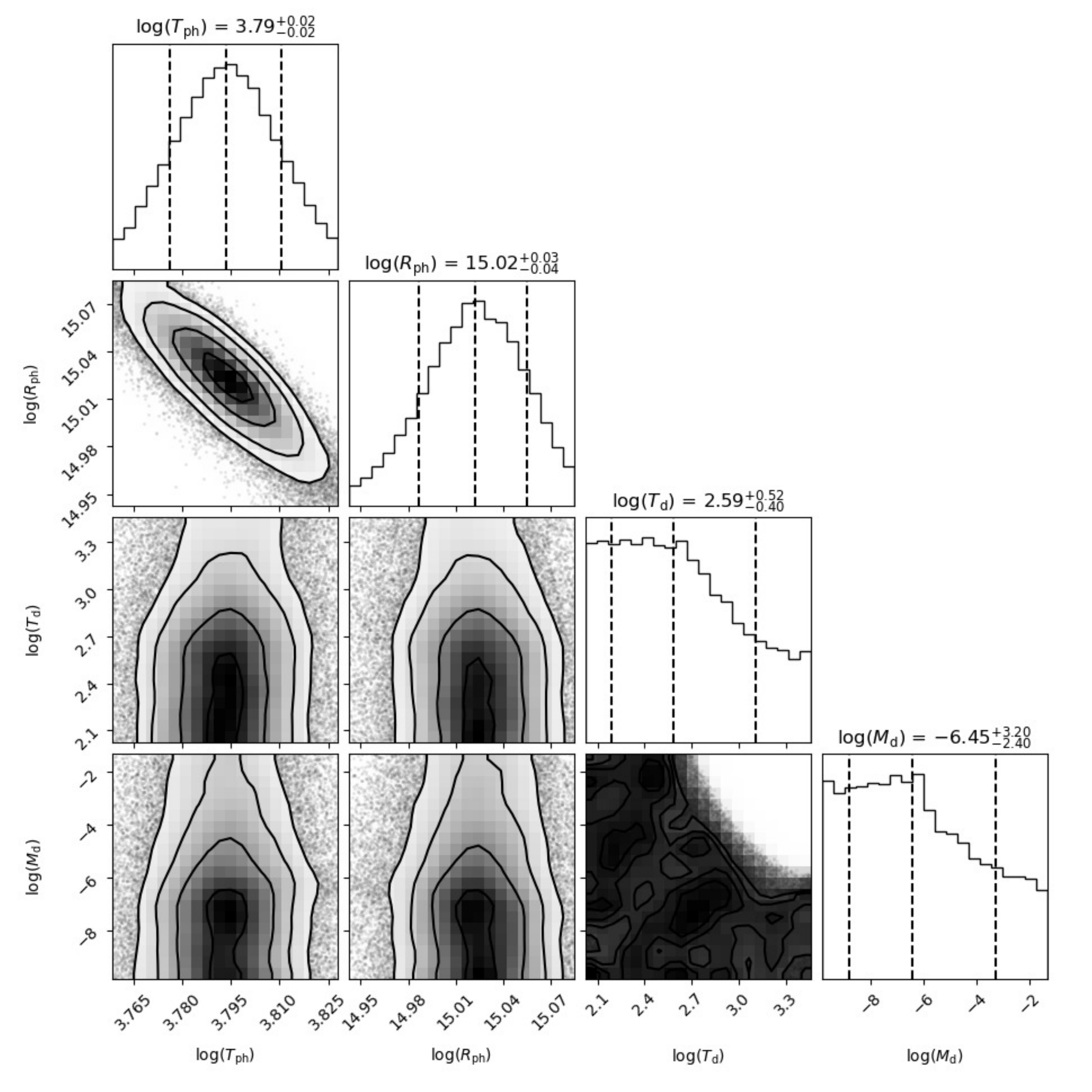}
\includegraphics[width=0.48\textwidth,angle=0]{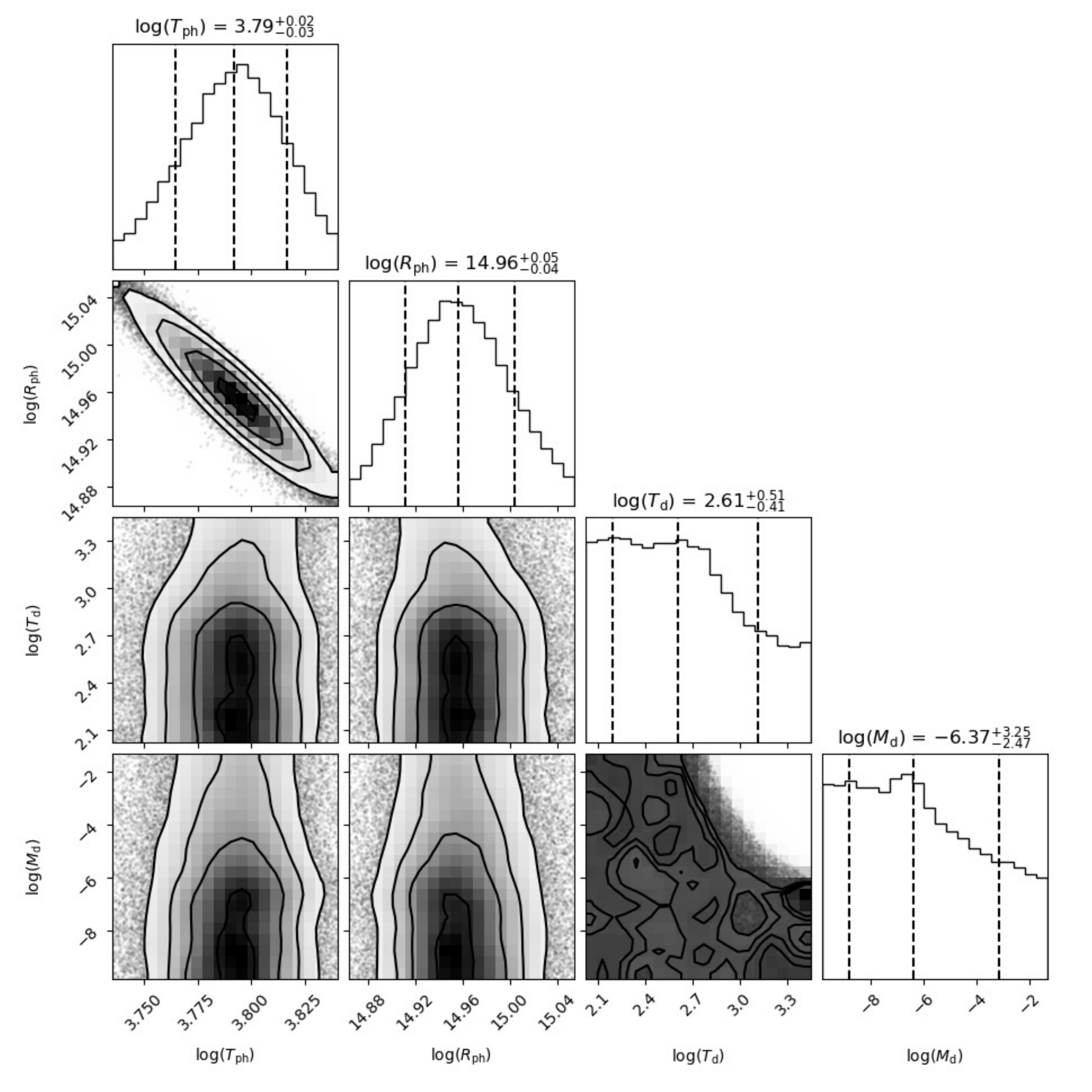}\\
\includegraphics[width=0.48\textwidth,angle=0]{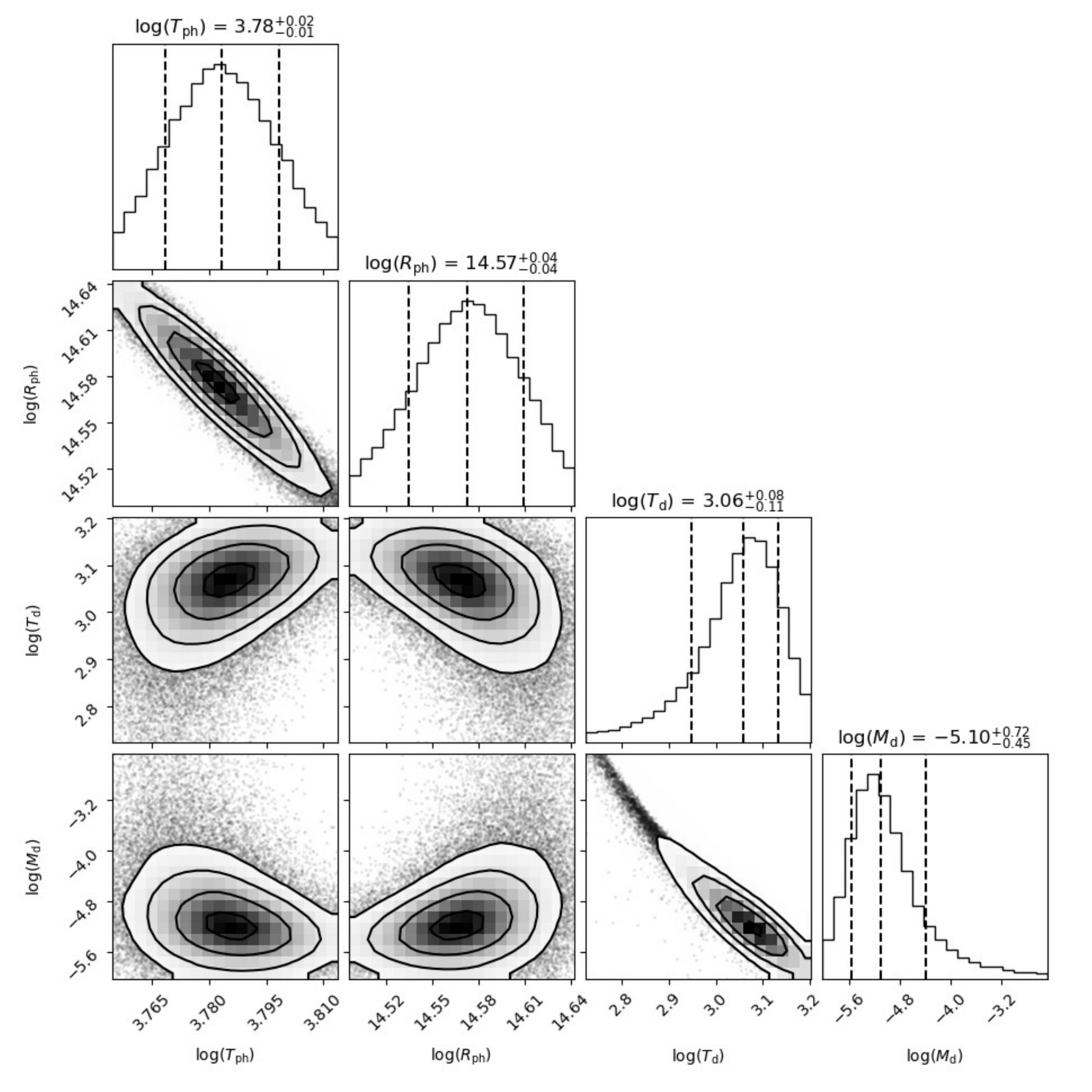}
\caption{The corner plots of the two-component model for SEDs of SN~2010al---continued.}
\end{figure}

\clearpage

\begin{figure}
\centering
\vspace{80pt}
\includegraphics[width=0.48\textwidth,angle=0]{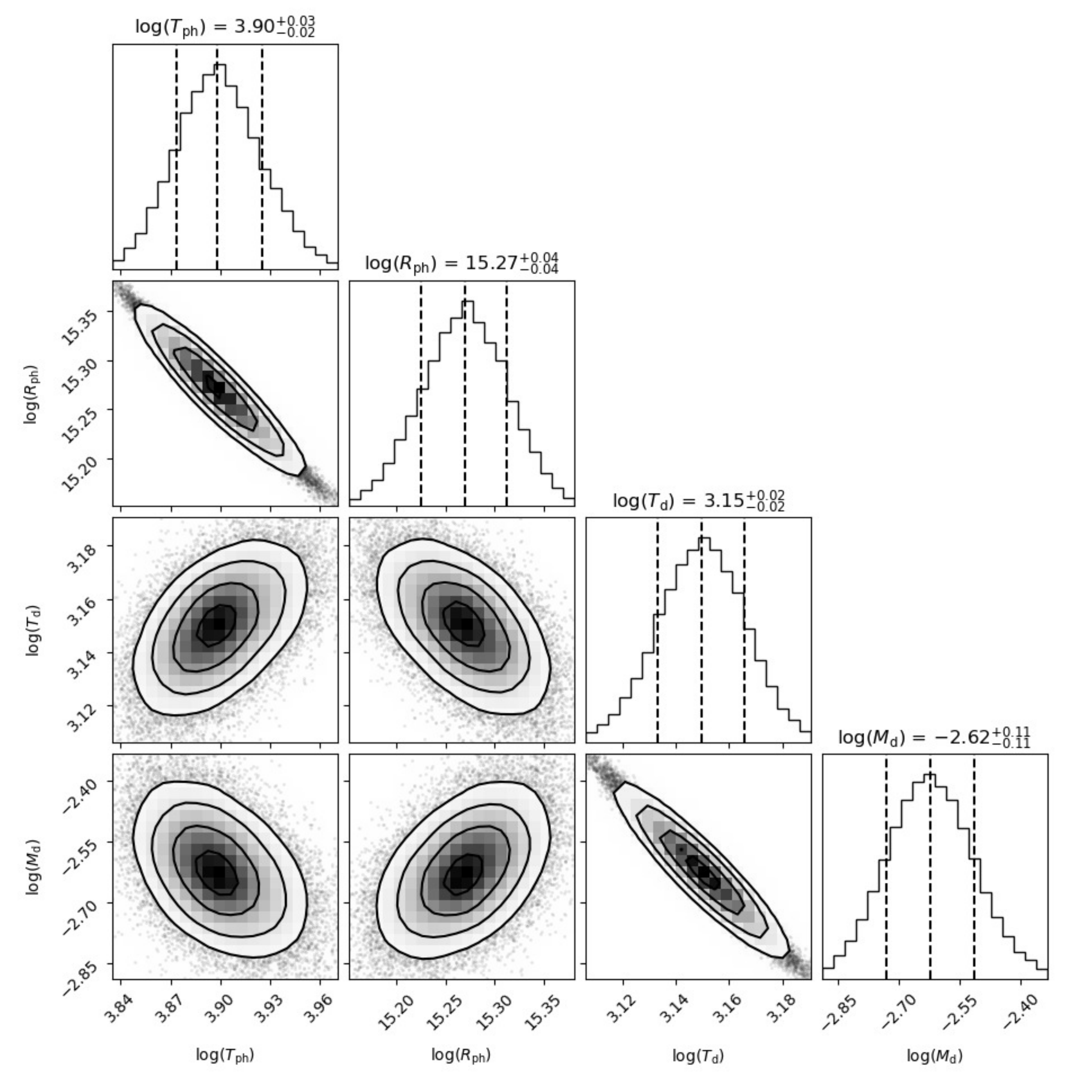}
\includegraphics[width=0.48\textwidth,angle=0]{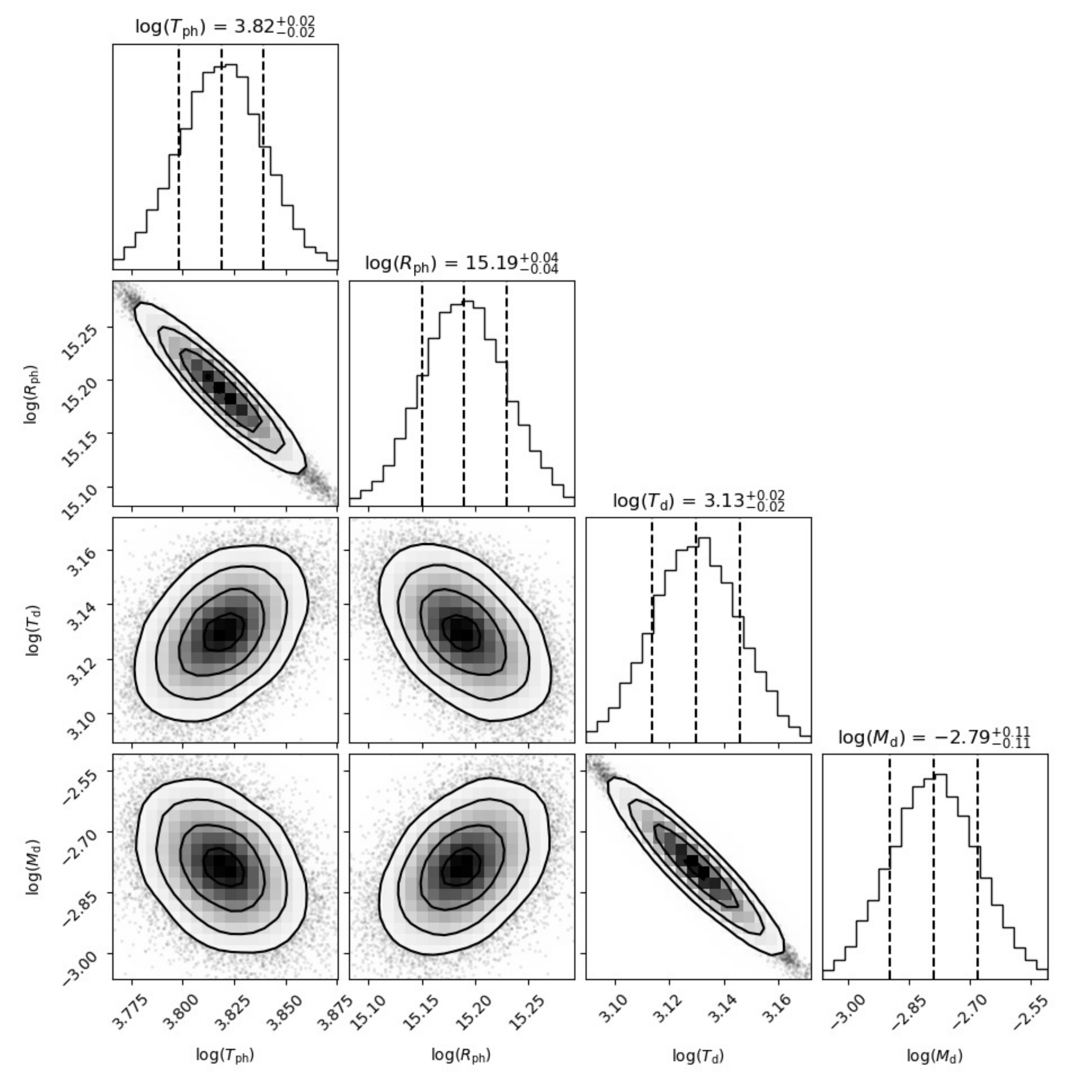}\\
\includegraphics[width=0.48\textwidth,angle=0]{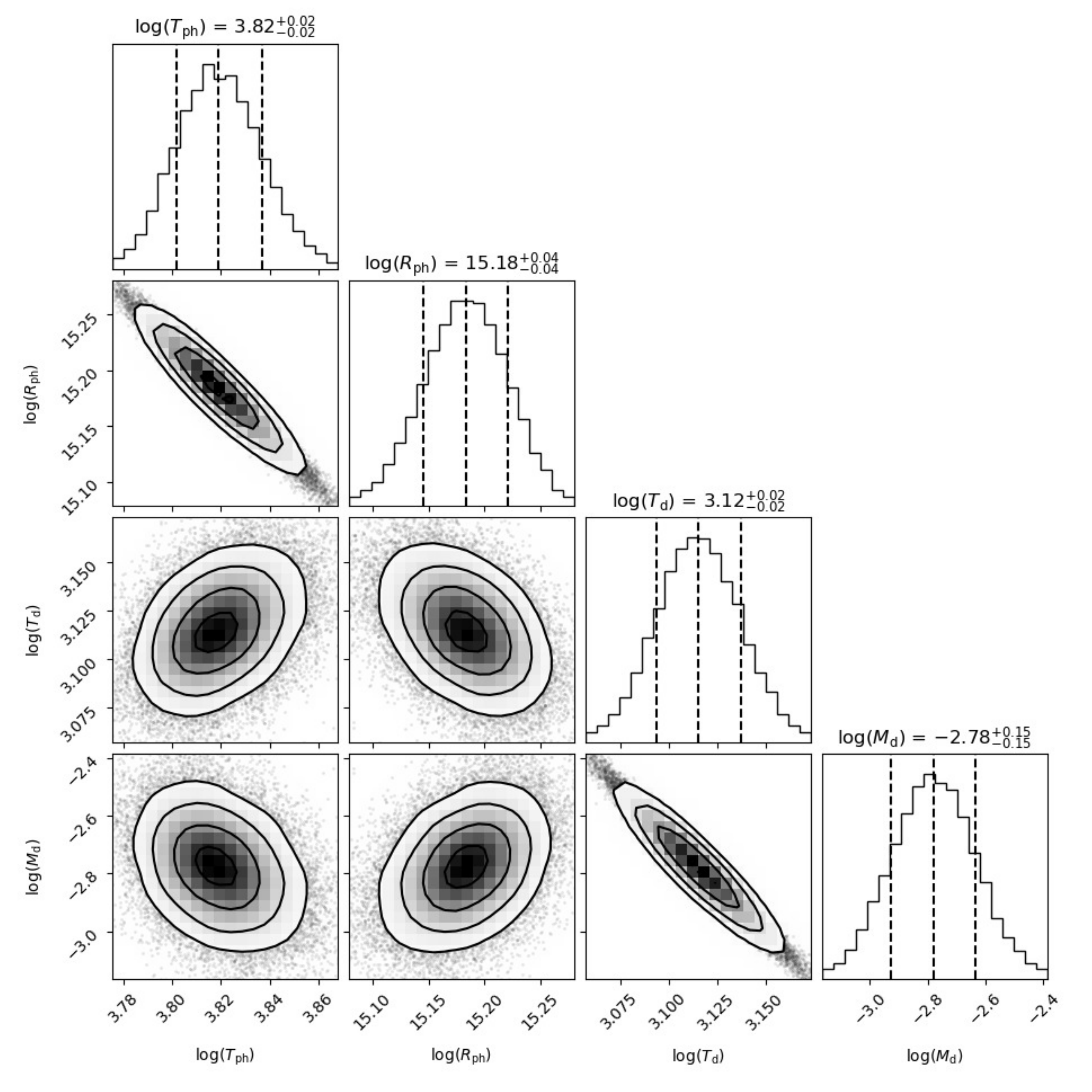}
\includegraphics[width=0.48\textwidth,angle=0]{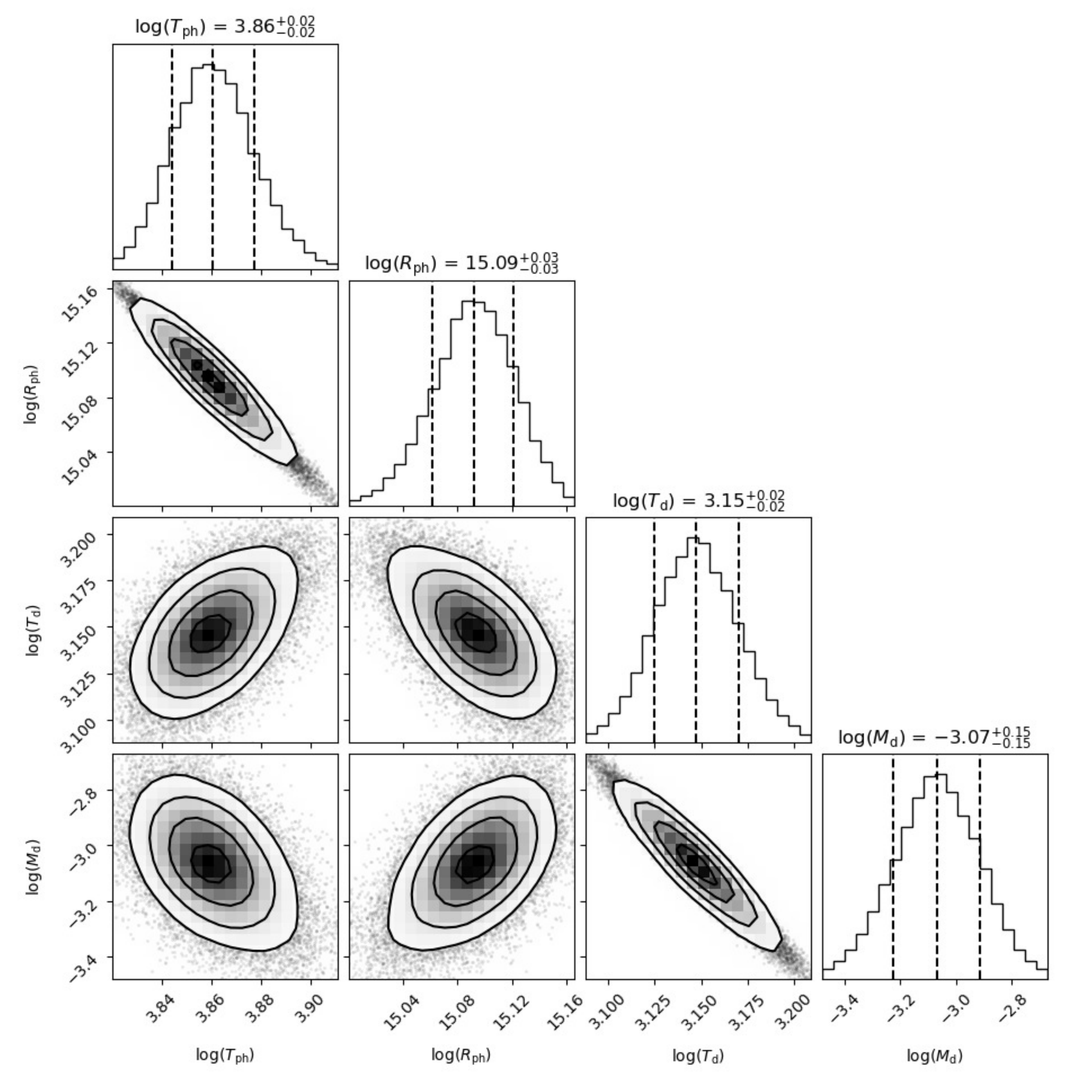}
\caption{The corner plots of the two-component model for SEDs of OGLE-2012-SN-006.}
\label{fig:2012corner}
\end{figure}

\begin{figure}
\ContinuedFloat
\vspace{80pt}
\centering
\includegraphics[width=0.48\textwidth,angle=0]{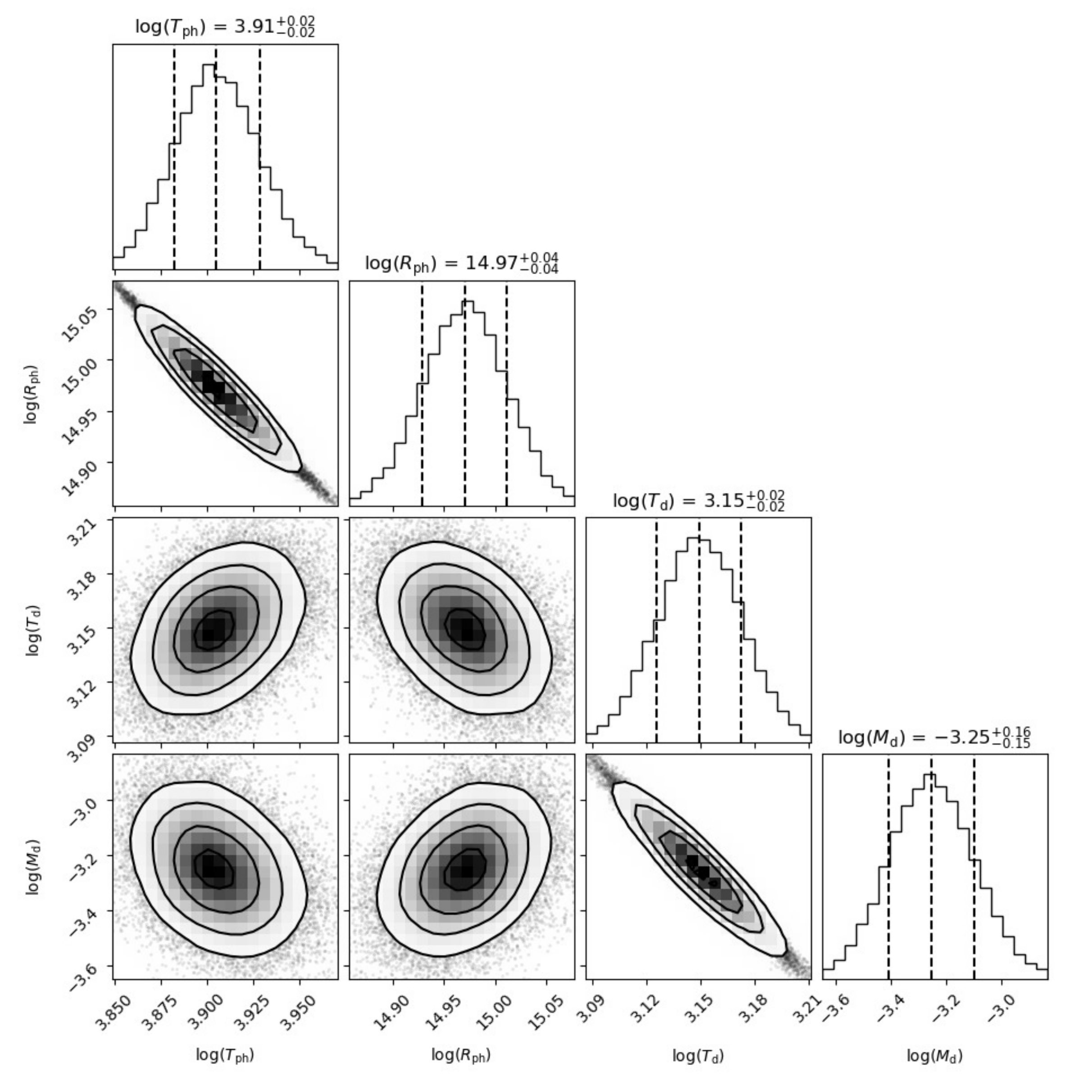}
\includegraphics[width=0.48\textwidth,angle=0]{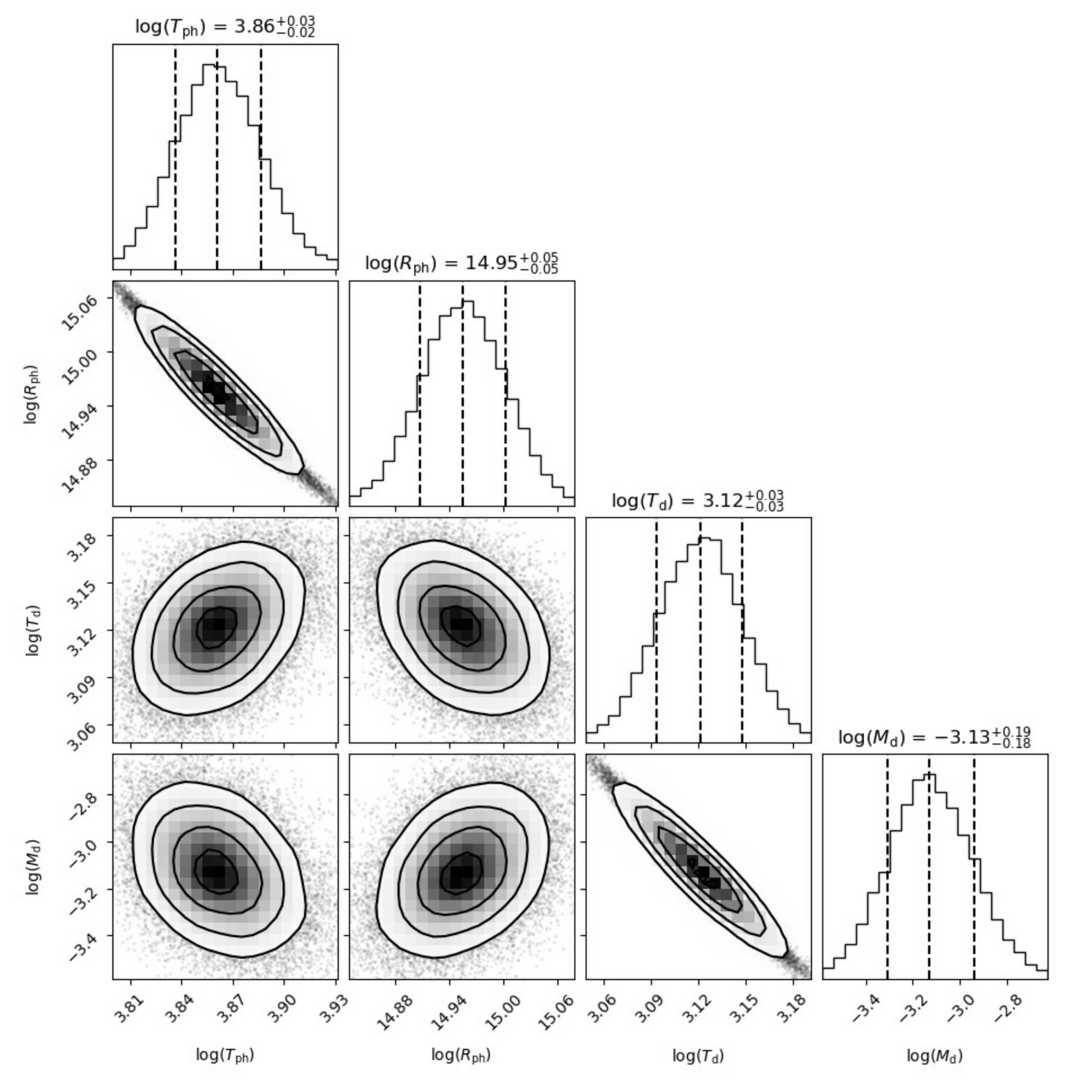}\\
\includegraphics[width=0.48\textwidth,angle=0]{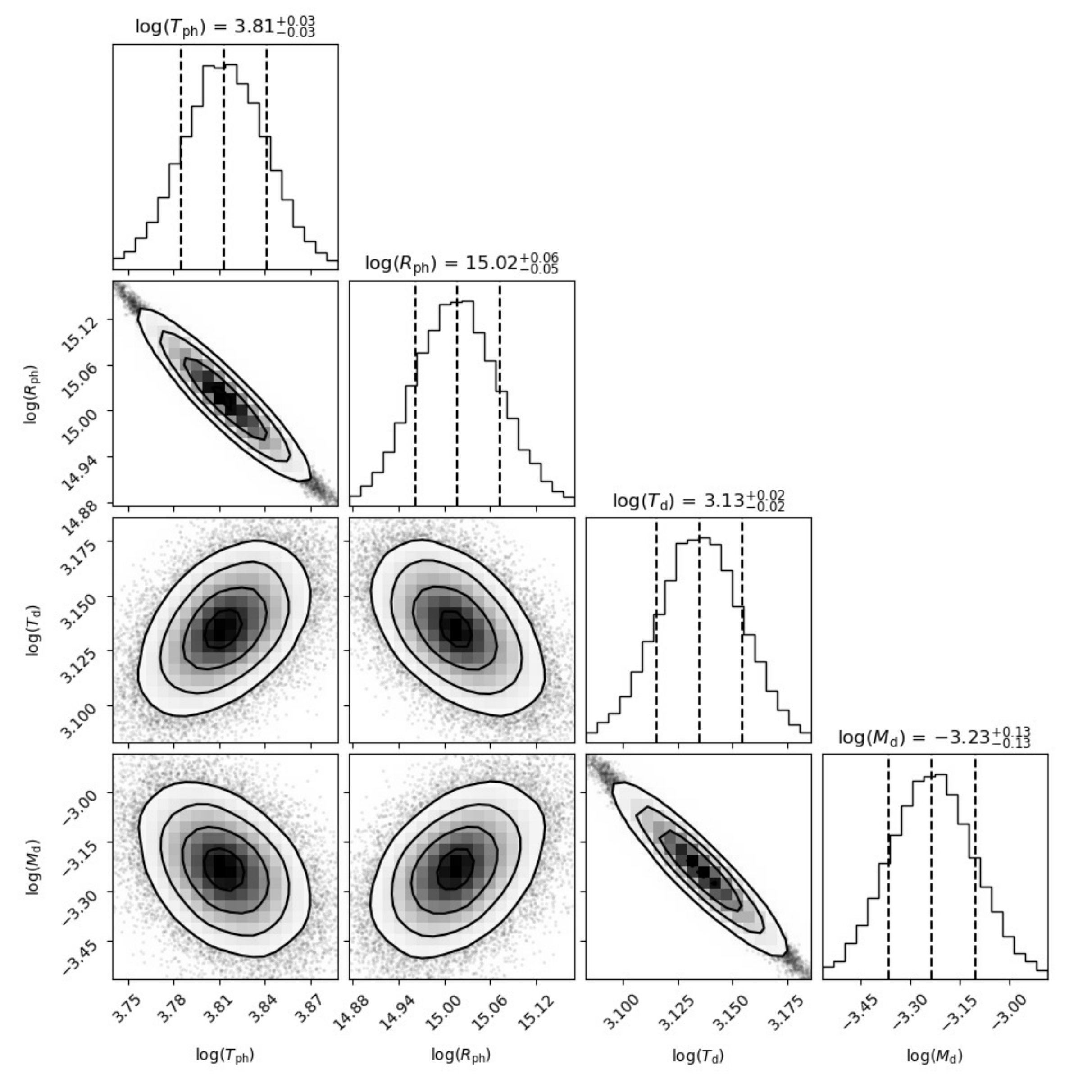}
\caption{The corner plots of the two-component model for SEDs of OGLE-2012-SN-006---continued.}
\end{figure}

\clearpage

\begin{figure}
\vspace{50pt}
\centering
\includegraphics[width=0.48\textwidth,angle=0]{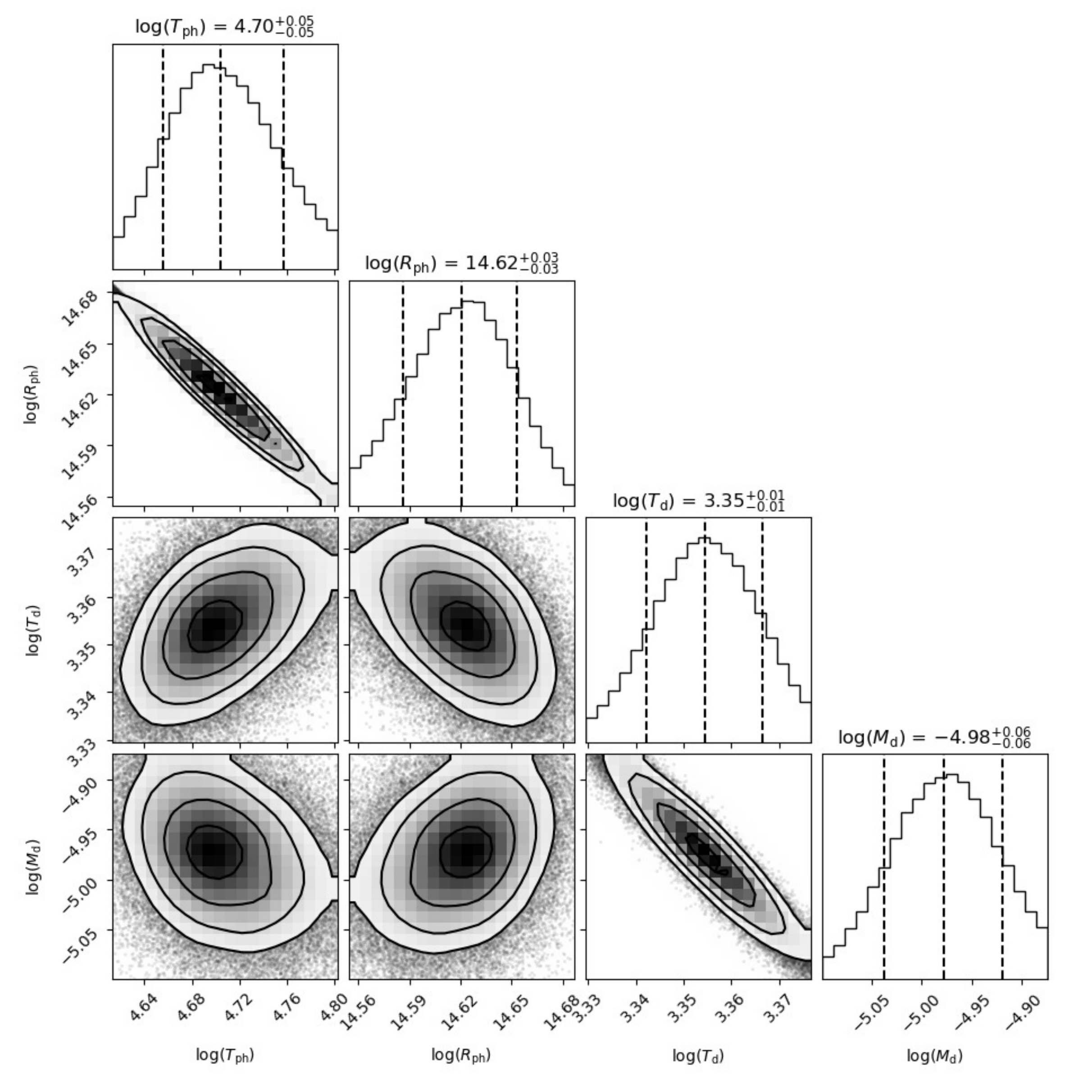}
\includegraphics[width=0.48\textwidth,angle=0]{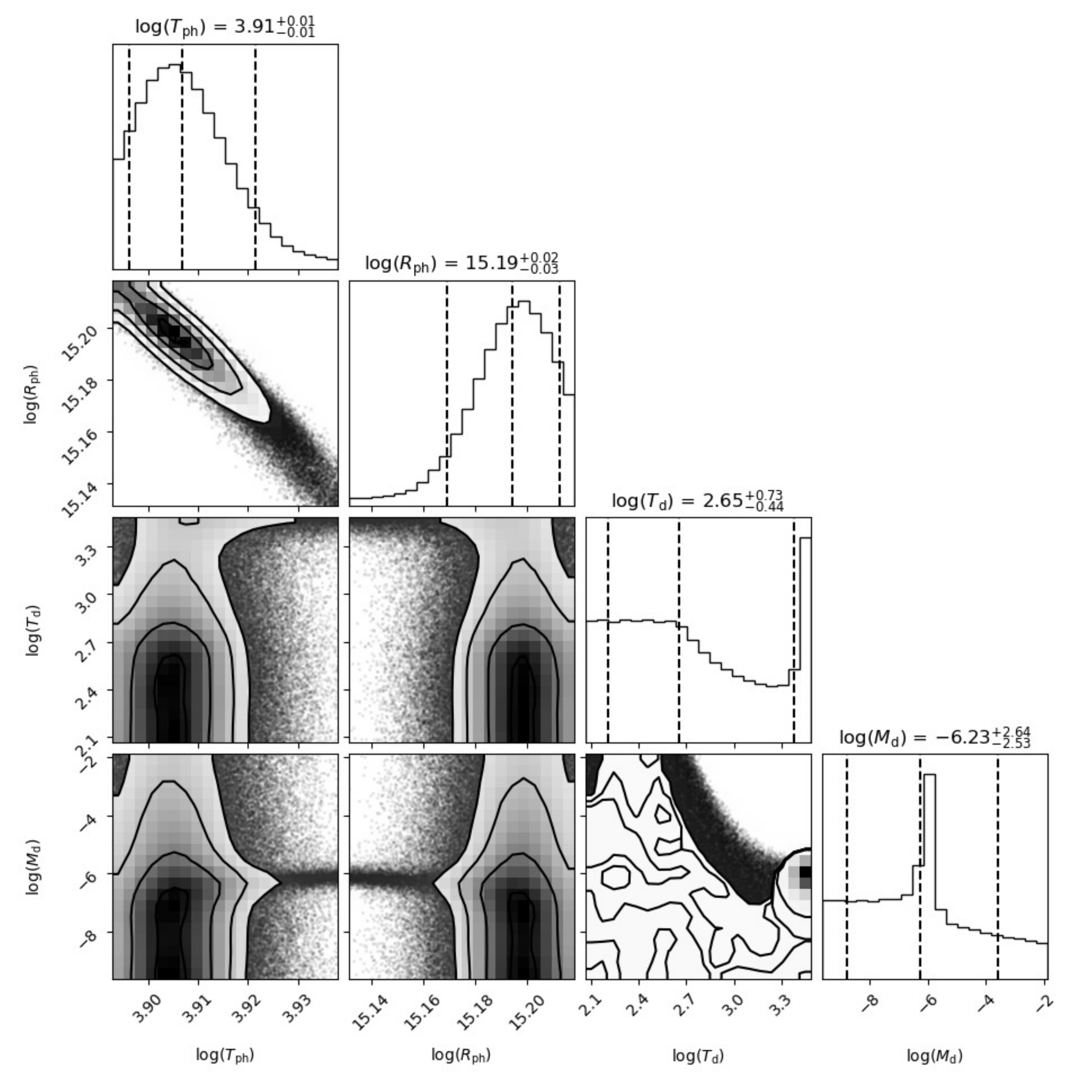}
\caption{The corner plots of the two-component model for SEDs of LSQ13ddu.}
\label{fig:2013corner}
\end{figure}

\begin{figure}
\vspace{-50pt}
\centering
\includegraphics[width=0.48\textwidth,angle=0]{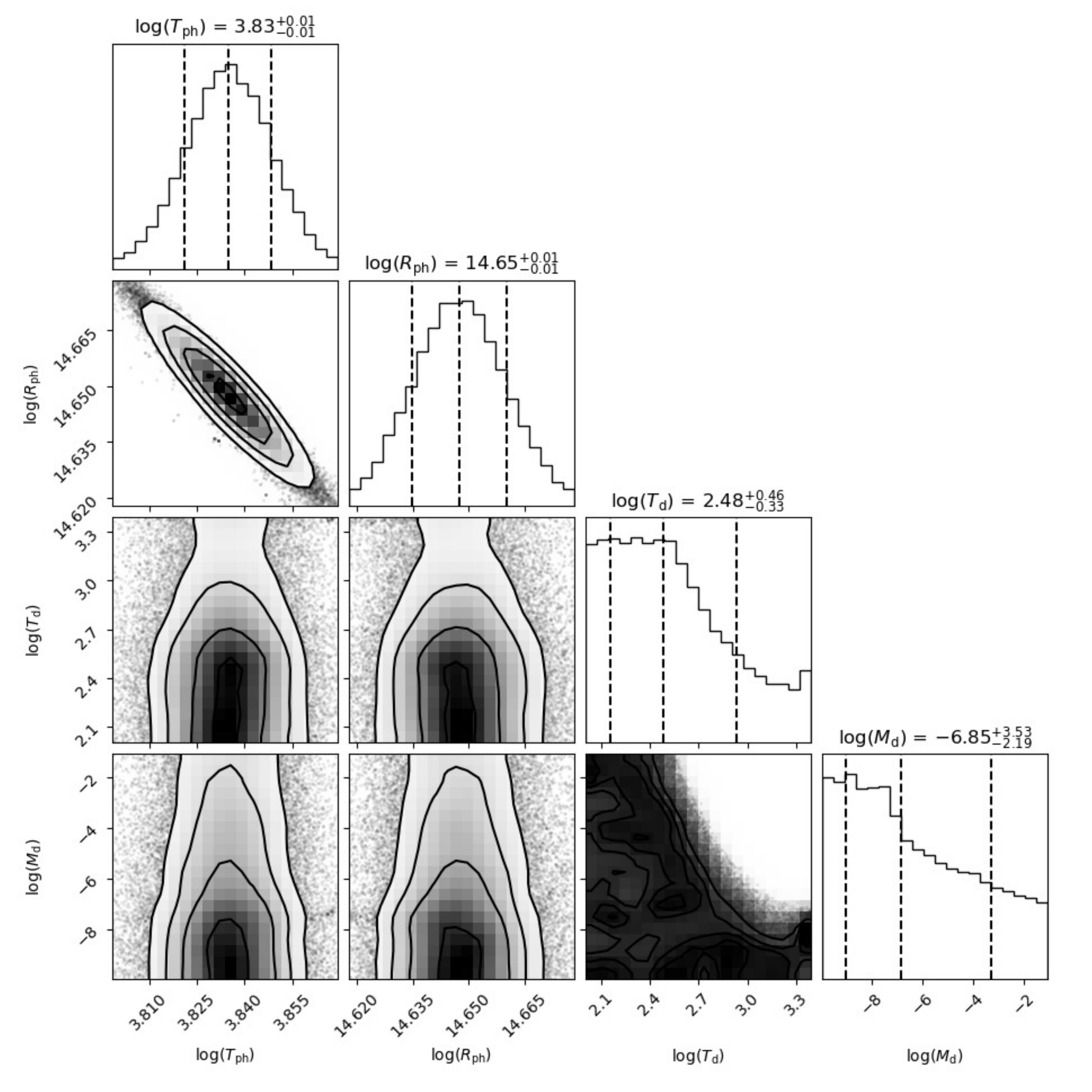}
\includegraphics[width=0.48\textwidth,angle=0]{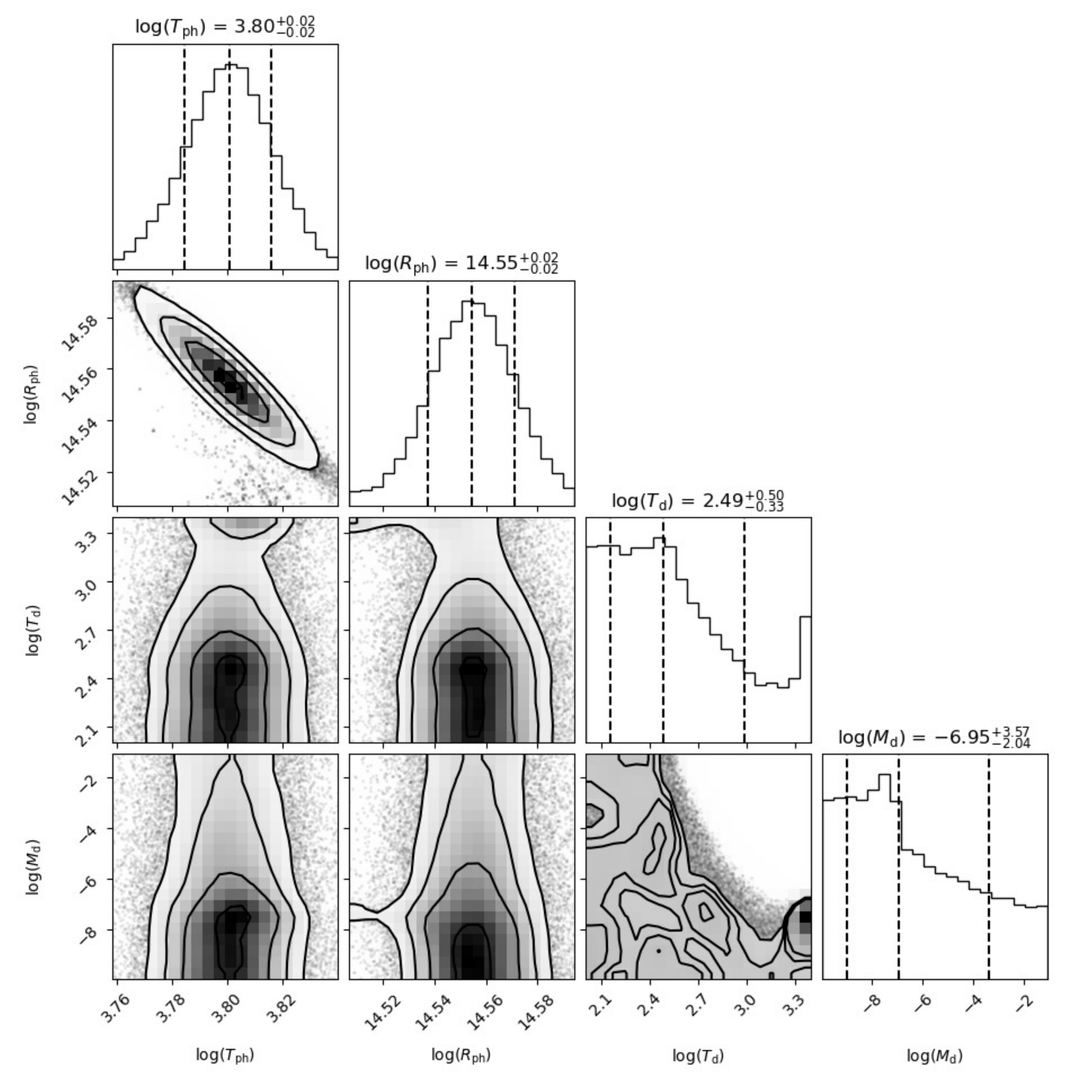}
\caption{The corner plots of the two-component model for SEDs of SN~2015G.}
\label{fig:2015corner}
\end{figure}

\end{document}